\documentclass[fleqn,usenatbib]{mnras}

\usepackage{times}
\usepackage[T1]{fontenc}
\usepackage{ae,aecompl}
\usepackage{graphicx}	
\usepackage{enumerate} 
\usepackage{balance}
\usepackage{amsmath}	
\usepackage{amssymb}	
\usepackage{chngcntr}
\counterwithout{table}{section}

\usepackage{array}
\newcolumntype{L}[1]{>{\raggedright\let\newline\\\arraybackslash\hspace{0pt}}m{#1}}
\newcolumntype{C}[1]{>{\centering\let\newline\\\arraybackslash\hspace{0pt}}m{#1}}
\newcolumntype{R}[1]{>{\raggedleft\let\newline\\\arraybackslash\hspace{0pt}}m{#1}}

\newcommand{\asec}{^{\prime\prime}}

\newcommand{\Mpc}{{\mathrm{Mpc}}}

\newcommand{\mpc} {\mathrm{Mpc}}

\title[SMF of galaxies, disks and spheroids at $z\sim0.1$]{Stellar mass functions of galaxies, disks and spheroids at $z\sim0.1$}

\author[Thanjavur et al.]{
Karun Thanjavur,$^{1}$\thanks{E-mail: karun@uvic.ca (KT)}
Luc Simard,$^{1,2}$
Asa F.L. Bluck$^{1,3}$
and Trevor Mendel$^{4}$
\\
$^{1}$Department of Physics \& Astronomy, University of Victoria, PO Box 1700, STN CSC, Victoria, BC, V8P 1A1, Canada. \\
$^{2}$NRC Herzberg, 5071 West Saanich Road, Victoria, BC, V9E 2E7, Canada. \\
$^{3}$Institute for Astronomy, ETH Zurich, Wolfgang-Pauli-Strasse 27, 8093, Zurich, Switzerland. \\
$^{4}$Max Planck Institute for Extraterrestrial Physics, Giessenbachstr. 1, 85748 Garching, Germany.
}

\date{Accepted XXX. Received YYY; in original form ZZZ}

\pubyear{2015}

\begin{document}
\label{firstpage}
\pagerange{\pageref{firstpage}--\pageref{lastpage}}
\maketitle


\begin{abstract}
We present the stellar mass functions (SMF) and mass densities of galaxies, and their spheroid and disk components in the local (z$\sim$0.1) universe over the range $8.9 \leq$ log($M/ M_{\odot}$) $\leq$ 12 from spheroid+disk decompositions and corresponding stellar masses of a sample of over 600,000 galaxies in the SDSS-DR7 spectroscopic sample. The galaxy SMF is well represented by a single Schechter function ($M^* = 11.116\pm0.011$, $\alpha = -1.145\pm0.008$), though with a hint of a steeper faint end slope. The corresponding stellar mass densities are ($2.670\pm0.110$), ($1.687\pm0.063$) and ($0.910\pm0.029$)$\times10^8 M_{\odot} \Mpc^{-3}$ for galaxies, spheroids and disks respectively. We identify a \textit{crossover stellar mass} of log$(M/M_{\odot})$ = 10.3$\pm$0.030 at which the spheroid and disk SMFs are equal. Relative contributions of four distinct spheroid/disk dominated sub-populations to the overall galaxy SMF are also presented. The mean disk-to-spheroid stellar mass ratio shows a five fold disk dominance at the low mass end, decreasing monotonically with a corresponding increase in the spheroidal fraction till the two are equal at a galaxy stellar mass, log$(M/M_{\odot})$=10.479$\pm$0.013; the dominance of spheroids then grows with increasing stellar mass. The relative numbers of composite disk and spheroid dominated galaxies show peaks in their distributions, perhaps indicative of a preferred galaxy mass. Our characterization of the low redshift galaxy population provides stringent constraints for numerical simulations to reproduce.
\end{abstract}

\begin{keywords}
galaxies: fundamental parameters  --- galaxies: mass function  --- galaxies: statistics  --- galaxies: evolution  --- galaxies: stellar content --- astronomical databases: miscellaneous
\end{keywords}

\section{Introduction} \label{intro}

\noindent Piecing together a comprehensive picture of the various complex and interconnected processes driving galaxy formation and evolution over the history of the Universe to the present times continues to be a major area of research in cosmology (see \citet{Cons14} for a recent review). Despite impressive advances in observational technology which permit resolving galaxies even within the first billion years of cosmic history at z$\geq6$ (\citet{Will13} for example), as well as on the theoretical front with detailed numerical simulations such as the \textit{Illustris} \citep{Voge14a}, the \textit{Eagles} \citep{Scha15} and \textit{FIRE} \citep{Hopk14} projects for example, many details regarding galaxy formation and evolution remain yet to be fully resolved and understood. 

\indent In the canonical model, galaxy formation may proceed by several parallel pathways including monolithic collapse followed by in situ star formation in the collapsed halo, by mergers between galaxies of differing mass ratios, and by the accretion of gas from the surrounding (intergalactic) medium into the potential well of a seed dark matter halo (\citet{Cons14} and extensive cross-references therein). The high angular momentum of the infalling gas during this formation process results primarily in a disk-dominated galaxy. Subsequent hierarchical evolution supports the picture that the merger of two such galaxies completely destroys the disks in the progenitors, and results in a spheroidal system \citep{Toom72, Barn92}. This simple evolutionary scenario leads to the extant `Hubble sequence' with the spheroids being the end products of disk-dominated progenitors. However, \citet{Sale12} present recent theoretical evidence that spheroids can form without mergers or disk instabilities when the gas infall is along filaments of `cold' flows leading to a spin misalignment between the infalling gas and the halo. Subsequent episodic star formation leads to stellar populations with radically different kinematics, which subsequently relax to the pressure-supported spheroids we observe today. 

\indent However, detailed observations \citep{Pipi14, Law12}, as well as recent high resolution N-body simulations have shown that this evolutionary picture may be too simple, and that a disk may also subsequently grow around this spheroidal merger remnant through the accretion of external cold gas \citep{Mitc14, Hopk13, Stew09a, Robe07, Bens02a}. The gas content of the merger progenitors plays a key role in shaping the characteristics of the resulting galaxy through a relatively rapid re-establishment of a disk component out of {\it in situ} gas around the central spheroid \citep{Hopk10, Stew09a, Spri05a, Burk04}. This important role of gas in determining the morphology of a merger remnant has led to the classification of mergers as \emph{gas rich} (``wet") and \emph{gas poor} (``dry"). As a result of such a series of mergers and gas accretions, the morphology of a galaxy may undergo dramatic transformations at various times in its life \citep{Stei02}. 

 \indent Intense theoretical efforts in recent years have been aimed at a better understanding of the mechanisms responsible for these disk-to-spheroid transformations, and for the build up of stellar mass in spheroids in galaxies of different Hubble types and of different masses \citep{Cons14, Aume14, Hopk13, Hopk12, Hopk10, Khoc11, Khoc09a, Khoc09b, Cox08, Korm04, Burk04, Burk03, Stei02, Cole00}. Various formation mechanisms have been proposed to be at work in transforming disks into spheroids, ranging from mergers (minor or major mergers depending on the mass ratio of the participants, gas rich or poor based on their gas content), to internal disk instabilities (secular evolution), and perturbations such as tidal interactions from neighboring galaxies \citep{Mitc14}. These theoretical models match observational results based on recent large galaxy surveys such as the SDSS-DR7 \citep{Abaz09}, PRIMUS (PRIsm MUlti-object Survey; \citet{Coil11,Cool13}) and GAMA (Galaxy And Mass Assembly; \citet{Driv09}) at least in broad principles though perhaps not in fine detail \citep{Jian12, deProp14}. 
 
 \indent The picture that is emerging from these efforts shows that characteristics of the spheroidal component of a galaxy at z=0 (bulge fraction, half light radius, relative stellar mass, $M/L$ ratio and age of the stellar population) are highly dependent on the merger history of the host galaxy \citep{Hopk09a}. More importantly, models also indicate that the merger history of a galaxy is a strong function of its stellar mass - a galaxy at the high end of the stellar mass function has a history which is very different, in a statistical sense, from that of a low mass galaxy \citep{Aume14, Hopk13, Naab07, Khoc06}. The number of major mergers as well as the redshift at which a galaxy experiences its first major merger are dependent on its mass - over 40\% of the population with stellar masses a few times $ M^*$ have undergone two or more major mergers at redshifts 2 or higher; on the other hand, nearly all the low mass galaxies with ($M \leq M^*$)\footnote{Throughout this manuscript, $M$ represents galaxy stellar mass, while $M^*$ is the characteristic stellar mass.} have their first, and perhaps only major merger below $z\sim1$ \citep{Hopk09a, Hopk09b, Hopk09c, Hopk09d, Hopk09e}; here, $M^* \sim 10^{11}M_\odot$ is the characteristic mass in the Schechter function representation \citep{Sche76} of the galaxy stellar mass function (SMF).  
 
 \indent Galaxies in the local universe are the end products of the evolutionary processes described by these theoretical models. Therefore, observed properties of the disks and spheroids of a statistically significant sample of low redshift galaxies are necessary to test and constrain these proposed methods for the assembly of stellar mass in these components. In pursuit of this objective, we present the SMF and the stellar mass densities of a galaxy population exceeding 600,000, as well as those of the spheroid and disk components of our entire sample. The significant sample size also permits us to subdivide the galaxies based on their stellar mass spheroid-to-total ratio ($B/T$) into spheroid- and disk-dominated populations, and thus obtain the SMF of these sub-groups (each exceeding $\sim$150,000 galaxies).  
 
\indent The availability of the stellar masses in the spheroid and disk components of each galaxy in this substantial, low-redshift sample of galaxies permits us to compute the SMF of the spheroidal component, and that of the disk component of galaxies. These are the SMFs of the two principal \textit{components} of the complete galaxy sample, and should be clearly distinguished from the SMFs of galaxy \textit{sub-populations}, which have been extensively discussed in literature, e.g. \citet{Kelv14a, Mous13, Bald12, Pozz10, Verg08}, and \citet{Bund06}. Other than being of interest in their own rights, the spheroid and disk SMFs also permit us to also compute an important quantity, which we term the \textit{crossover stellar mass} of these galaxy components (again to be distinguished from a crossover mass for the galaxy sub-populations discussed in \citet{Bolz10, Pozz10, Verg08} and others). The SMF of the spheroid and disk components show distinct regions of spheroid and disk dominance. Between these regions lies a stellar mass where the spheroidal and disk SMFs intersect and are thus equal. This \textit{crossover stellar mass} may provide a much stronger observational constraint to simulations on galaxy evolution than just each SMF taken individually. The validity of any galaxy evolution model, which aims to capture all the complexities of both the spheroid and disk growth, depends not just on reproducing the shape and normalization of the individual SMF, but should also capture this crossover stellar mass between the two galaxy components.
 
 \indent Our results are based on the public \emph{GIM2D} catalog (\citet{Sima11}, hereafter $S11$) containing photometric (in the SDSS $g$ and $r$ filters) and structural properties of 603,122 galaxies selected from the Legacy area of the Sloan Digital Sky Survey Data Release Seven (\emph{SDSS-DR7}). The corresponding stellar masses for this entire galaxy sample and the spheroid and disks have been computed with SED fitting to all five SDSS wavebands by \citet{Mend14} (henceforth $MT14$) and are also available as part of their public release catalogs. In an accompanying publication (Thanjavur et al. 2016, \textit{in preparation}), we also present the corresponding five band $ugriz$ luminosity functions of this galaxy sample and of their disk and spheroid components. 
 
 \indent The manuscript is organized as follows: in Section \ref{Cats} we describe the $S11$ GIM2D (\S \ref{GIMCats}) and $M14$ stellar mass catalogs (\S \ref{SMCats}) as well as the selection criteria we apply to obtain our galaxy sample. A description of the galaxy sub-classes we use are provided in \S \ref{BDclasses}, and pertinent characteristics of the galaxy population in our catalogs, various corrections made to ensure consistency of the stellar masses may be found in \S \ref{CatProps}. We estimate the completeness of our catalogs in \S \ref{mur_comp} with the bivariate r-band surface brightness and stellar mass distribution. \S \ref{VmaxCorr} details the $1/V_{max}$ completeness corrections we consequently apply. We use these well characterized galaxy catalogs to compare the correlations between galaxy color to stellar mass distribution of the disk- and spheroid-dominated populations in \S \ref{ClrMig}. We begin our results in Section \ref{Results} with the SMF of the complete galaxy sample (\S \ref{SMFg}) followed by a comparison of our results with other recent published values (\S \ref{ExtcompSMFg}).  \S \ref{SMFbd} gives the stellar mass functions of their spheroid and disk components of the complete galaxy sample, as well as their \textit{crossover stellar mass}. We present the SMFs of the four galaxy sub-populations defined by their $B/T$ ratios in \S \ref{SMFbddom}, with a comparison with other published results in \S \ref{ExtcompSMFds}. The aim is to attempt to characterize the effects of differing merger histories on the SMF of these morphologically distinct populations. The complementary question of how much spheroids and disks contribute respectively to the stellar mass budget of the local galaxy population is addressed next in \S \ref{bd2gSM}; in order to verify any dependency of their relative contributions on the stellar mass of the host galaxy, this comparison is also carried out independently for our four morphology bins. Finally, in \S \ref{Nbd2gSM}, we present a census of the relative numbers of spheroid- and disk-dominated galaxies in stellar mass bins. We discuss our findings and the implications of the comparisons with existing literature in Section \ref{Disc}. In the final Section \ref{Sum}, we provide a summary of our principal results and findings. In the Appendix \ref{LSS} we explore any impact of the presence of large scale structure in the SDSS survey region on the SMFs we have obtained. In the Appendices \ref{IntcompSMFg} - \ref{SystErrChksSMFg} we describe our test results from the internal consistency checks for the stellar masses in our catalogs, as well as those in the $MT14$ stellar mass catalogs due to the modelling assumptions made. 
 
 \indent Stellar masses used in our work have been computed using the \citet{Chab03} IMF. The SDSS photometry has been corrected with $k$-corrections by \citet{Blan07} and the \citet{Calz00} extinction law. The cosmology adopted throughout this paper is ($H_0, \Omega_m, \Omega_\lambda$) = (70 $km.s^{-1}.\mpc^{-1}$, 0.3, 0.7).

\section{Catalogs} \label{Cats}
\subsection {GIM2D catalogs of SDSS DR7 galaxies} \label{GIMCats}

\indent  We present the stellar mass function (SMF) of a low redshift ($z\sim0.1$) galaxy population of over 600,000 galaxies drawn from the SDSS-DR7 using their stellar masses available in the public release of the \citet{Mend14} ($MT14$) catalogs\footnote{http://vizier.cfa.harvard.edu/viz-bin/VizieR?-source=J/ApJS/210/3}. The stellar masses in these catalogs were estimated by SED fitting to the five colour SDSS $ugriz$ photometry of the galaxies. We also present the SMF of the disk and spheroidal components of these galaxies using their stellar masses which are also available in the $MT14$ catalogs. The structural and photometric properties of these components needed for the stellar mass estimates are drawn from the spheroid+disk decompositions of the galaxies available in the $SL11$ public catalogs\footnote{http://cdsarc.u-strasbg.fr/viz-bin/Cat?J/ApJS/196/11}. Complete descriptions of the steps used in building both these catalogs are given in $SL11$ and $MT14$ respectively; here we summarize only pertinent details. 

\indent GIM2D \citep{Sima02} is a dedicated galaxy decomposition software which deblends galaxies from the sky background and neighboring objects in single filter imaging. It then fits an exponential disk and a de Vaucouleurs profile \citep{deVau48} to the light distribution of the deblended object, and thus identifies the disk and spheroidal components respectively. Details of the methodology used for the deblending and decomposition processes, as well as the resulting set of rest frame properties computed by GIM2D are given in \citet{Sima02}. The selection function and other performance characteristics of GIM2D have been extensively tested and are reported elsewhere \citep{Sima02, Tasc11}. The current version of GIM2D (version 3.2) has the capability to simultaneously fit the structural parameters in more than one filter, as was the case for the public SDSS-DR7 GIM2D catalogs presented in $SL11$; the interested reader is referred to their detailed discussion on the advantages of this approach of simultaneously fitting in two filters. The photometric properties used by $MT14$ for the SED fitting were obtained by simultaneously fitting the $r$ filter, taken to be the fiducial waveband, with each of the $ugiz$ filters as described in \citet{Sima02} and $SL11$.

\indent $SL11$ also describe the tests carried out on various methods for background estimation and deblending, and the advantages of the ones used to generate the final version of the catalogs. In addition to the two-component spheroid+disk decomposition, the current GIM2D version also fits a single-component pure S\'{e}rsic profile \citep{Sers63} to each galaxy. As discussed by $SL11$, fitting a more complex spheroid+disk model to a galaxy whose brightness distribution is well described by a single-component profile may lead to fitting a spurious component due to the extra degrees of freedom. $SL11$ use the $F$-statistic (e.g. \citet{Bran89}), denoted in the GIM2D catalogs by the parameter $P_{pS}$ (\textit{probability of a pure S\'{e}rsic}), as a quantitative criterion to determine whether a two component model is needed. The interested reader is referred to $SL11$ for a complete discussion of the $P_{pS}$ thresholds used, and to the findings of \citet{Bluc14} for issues related to the $P_{pS}$ parameter and its effect on the estimate of the B/T stellar mass ratio. 

\indent Following this morphological decomposition, an extensive set of structural and photometric properties of the fitted spheroid/disk component(s) as well as the corresponding properties of their parent galaxies are computed and written to catalogs. For each galaxy, the GIM2D processing flag, $prcflag$ is used to report the reduced $\chi^2_\nu$ goodness-of-fits to the structural models, with the value of $prcflag=0$ indicating a successful fit in \textit{both} filters. For the photometry in each band, the total flux is obtained by integrating each fitted component from r=0 to infinity. As explained in $S11$, appropriate $k$-corrections \citep{Blan07} have been applied in computing all the photometric values in the GIM2D catalogs, which have also been corrected for Galactic extinction using values taken from the SDSS database. The GIM2D catalogs for the SDSS-DR7 release contain only extended objects (SDSS morphological type = 3), which have also been flagged by SDSS as being unsaturated and properly deblended. 
 
\begin{table*}
\caption{Classification of galaxy sub-groups}
 \begin{center}
 \scriptsize
 {\renewcommand{\arraystretch}{2.0}
 \begin{tabular}{|L{3cm}|C{2cm}|C{2cm}|C{2cm}|C{2cm}|}
 \hline
Class & $(B/T)_{lower}$ & $(B/T)_{upper}$ & Number \\
 \hline
 \hline
  BT100 & 0.8 & 1.0 & 199696 \\
 \hline
BT80 & 0.5 & 0.8 & 156797 \\
 \hline
 BT50 & 0.2 & 0.5 & 152007 \\
 \hline
 BT20 & 0. & 0.2 & 94622 \\
 \hline
 \end{tabular}
 }
\end{center}
\begin{flushleft} 
{\footnotesize Our adopted classification of galaxy sub-groups based on their stellar mass $(B/T)_*$ values. Listed are the names of the classes, lower and upper $(B/T)_*$ limits used for their definition, and the number of galaxies in each sub-group}
\end{flushleft}
\label{Tbl1}
\end{table*}
\normalsize

\subsection {Stellar mass catalogs of SDSS DR7 galaxies} \label{SMCats}

\indent The role of stellar mass as the defining characteristic of galaxies has been highlighted by several complementary studies in literature \citep{Guid15,Bluc14,Kelv14b,LiMao13,Mous13,Disn08,Cons06}\footnote{Given the extensive literature available, this list is admittedly very incomplete}. In particular, the stellar mass has been observed to closely correlate with other principal properties such as galaxy color \citep{Bald08, Cons06}, metallicity \citep{Moro15,Trem04}, and the observed effective radius \citep{Pate13,Dokk10}. $MT14$ use the five-filter broadband photometry available in the GIM2D catalogs for the galaxies and their spheroidal and disk components to estimate the corresponding stellar masses by SED fitting to a library of synthetic stellar populations. Adopting a \citet{Chab03} IMF, these synthetic stellar populations have been generated to span a realistic range of population age, metallicity, attenuation due to dust (modeled with the \citet{Calz00} law), as well as a smoothly declining star-formation rate and history. $MT14$, Table \ref{Tbl2a} lists the ranges of the various parameters used to generate this grid of stellar populations. 

\indent Galaxy stellar mass is not a directly observable quantity but is inferred from other measurements, therefore uncertainties in these estimates due to the adopted methodology need to be assessed. The SED fitting procedure used by $MT14$ is based on the \textit{flexible stellar population synthesis} ($FSPS$) method \citep{Conr10a}. $MT14$ report an overall \textit{statistical} uncertainty of $\leq$0.1 dex in the galaxy, spheroid and disk stellar masses given in their catalogs. More importantly, they carry out a detailed analysis of the corresponding \textit{systematic} uncertainties due to various assumptions inherent in the SED fitting. By their assessment, the combination of assumptions made toward the synthesis model, mainly the age and evolution of the stellar population, the initial mass function and the extinction law used may lead to corresponding changes in the stellar masses from 0.1 to 0.2 dex ($\sim25-60\%$); for an in depth discussion on the systematic uncertainties inherent in $FSPS$, refer to \citet{Conr09, Conr10b}, and also to \S4 of $MT14$. 

\indent On a related note, \citet{Bern13} strongly caution about the significant systematic uncertainties arising out of the differing light profile fits, and the assumptions made in the light-to-mass ratio conversions and SED fitting, \textit{especially at the high mass end of the SMF}. Even though the size of SDSS galaxy population minimizes the statistical uncertainties in the global SMF results we present, following the cautionary notes cited above regarding systematic uncertainties in stellar mass estimates, we accept that these inherent uncertainties in the $MT14$ catalogs will propagate through to our SMF estimates. Therefore, we attempt to quantify the effect of these systematic uncertainties in the $MT14$ stellar masses on our SMF results by using Monte-Carlo error estimations, as described in \S \ref{SMFbd}.


\begin{figure*} 
\centering
\includegraphics[width=0.99\columnwidth]{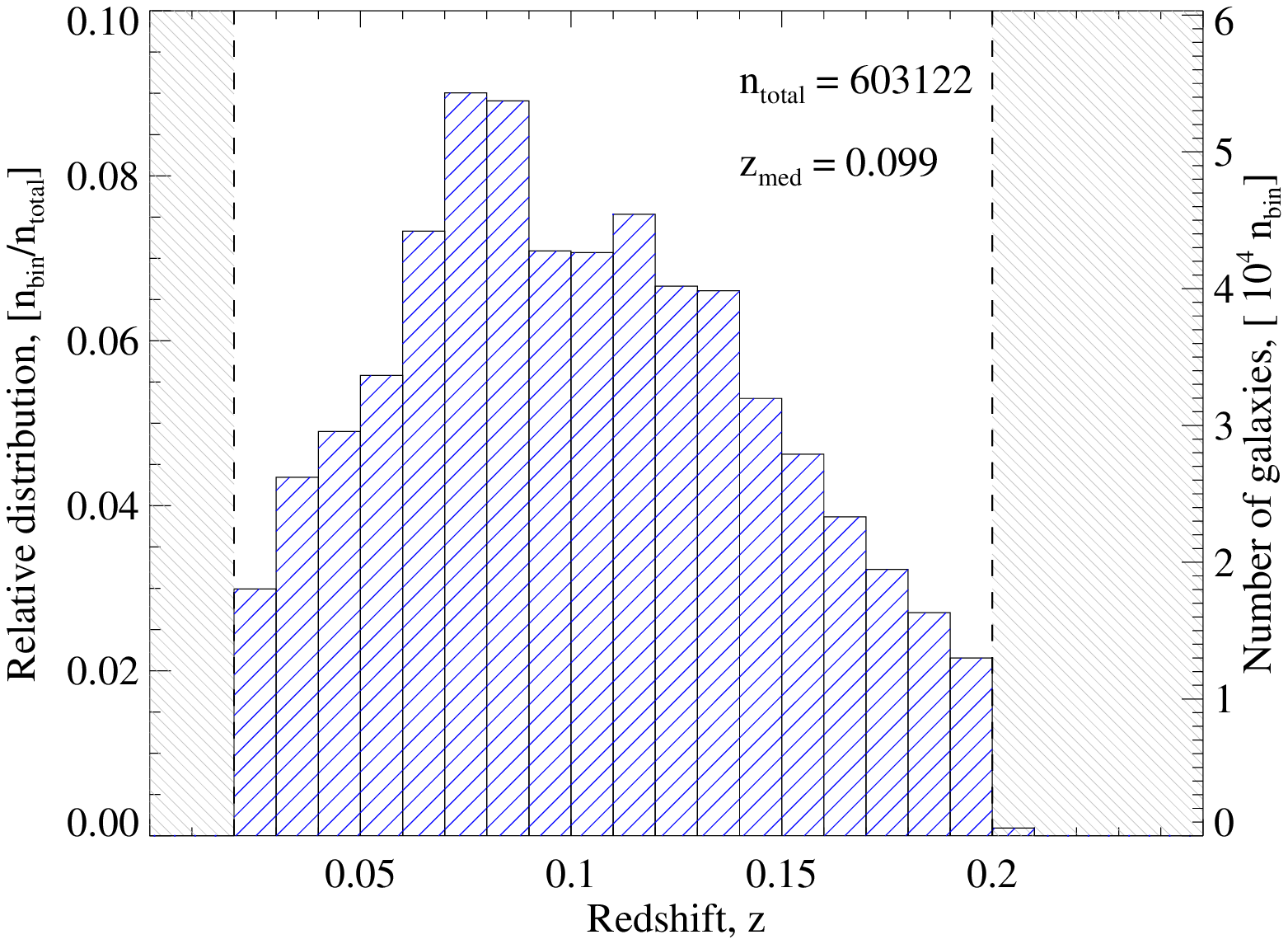}
\includegraphics[width=0.99\columnwidth]{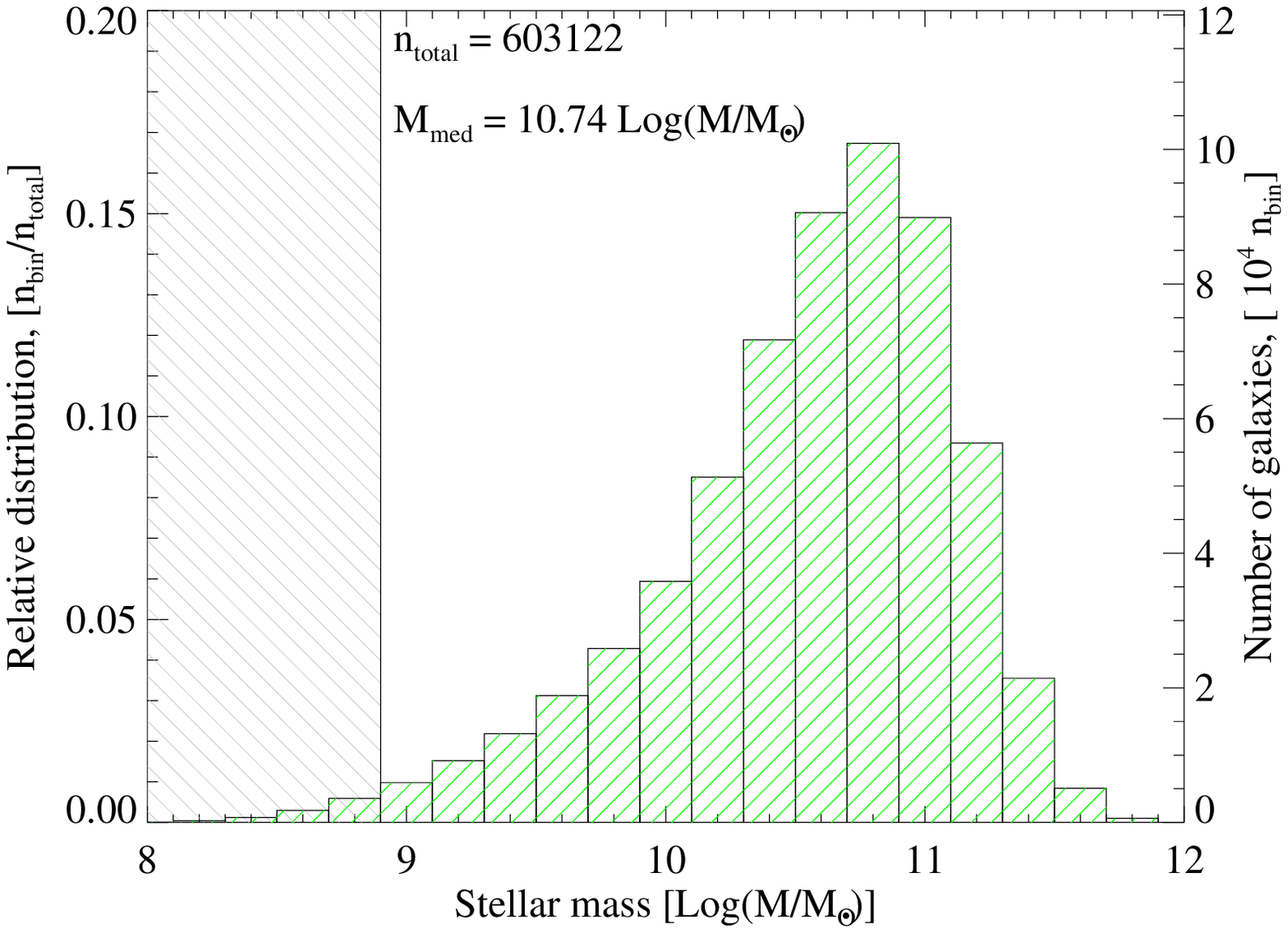}
\caption{(\textit{Left}) Distribution of the galaxy population as a function of redshift in the final catalog selected using the criteria described in \S \ref{Cats}; the hatched regions demarcate the spectroscopic redshift range we have adopted for our galaxy sample. The y-axes are scaled for both the fractional distribution (relative to the total number of galaxies), as well as the actual number of galaxies in each z=0.01 bin. (\textit{Right}) Histogram of galaxy stellar masses for the SDSS population in our catalog, with the y-axes scaled similar to the redshift histogram. The distribution shows a few $10^4$ galaxies per stellar mass bin, which thus provide a robust estimate of the SMF. (\textit{Color figure available in the online journal}) \label{Fig1}}
\end{figure*}

\indent In order to obtain a well characterized galaxy population we apply the following additional selection criteria: we only select galaxies which have also been observed as part of the SDSS spectroscopic survey and which have confirmed redshifts (SDSS $specclass$ = 2); spectroscopic redshifts are needed to compute the $1/V_{max}$ completeness correction (\citet{Schm68}, \citet{Felt76}), which we apply to our galaxy sample. By selecting only galaxies from the SDSS spectroscopic survey, we automatically inherit the corresponding photometric limits: surface brightness, $\mu_r \leq 23\,\mathrm{mag.arc.sec^{-2}}$, and r-band Petrosian magnitude limits, $14\leq r_{petro} \leq 17.77$ mag \citep{Stra02}. Based on the SDSS spectroscopic completeness criterion, we restrict the redshift range to $0.02 \leq z \leq 0.2$. We only select galaxies for which the GIM2D morphology fit has been flagged as successful (GIM2D $prcflag$ = 0) in \emph{all} filters. Finally, from the $MT14$ stellar-mass catalogs, we only select objects whose photometry is flagged as uncontaminated from nearby neighbours (\textit{contam\_flag}=0, and -0.2$\leq$ \textit{delta\_fiber} $\leq$ 0.2). The combination of the selection cuts for GIM2D $prcflag$ flag and photometric contamination in the $MT14$ catalogs removes only 4752 galaxies, ($\sim0.8\%$ of the total sample) with no discernible bias toward any stellar mass. 

\indent Based on the combination of all these selection criteria, we obtain a well characterized dataset for 603,122 galaxies, as well as for their spheroidal and disk components, with photometric and structural properties from the $SL11$ GIM2D public catalogs, and the $MT14$ catalogs providing their stellar masses. 

\begin{figure*} 
\includegraphics[scale=0.7]{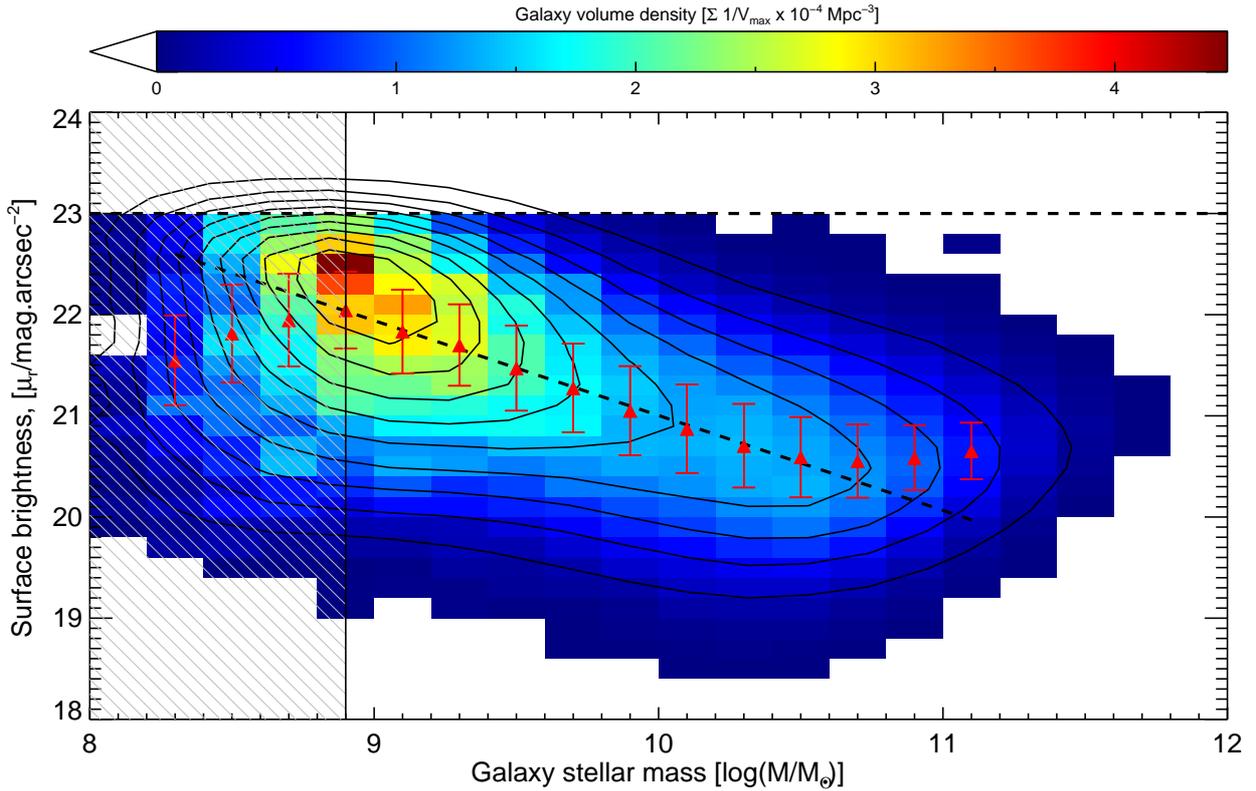}
\caption{ The correlation between galaxy stellar mass, $log(M/M_{\odot})$  and $r$-band surface brightness, $\mu_r$ plotted as the $1/V_{max}$ weighted \textit{bivariate brightness distribution} (BBD, \citet{Cros01}). The contours represent ten equally spaced intervals in the galaxy volume density in the range 0 to 4.47$\times 10^4 Mpc^{-3}$, shown by the colorbar. Overplotted is the trace of the central value and 1$\sigma$ width of gaussians fitted to the BBD in bins of 0.2 dex in stellar mass for the stellar mass range $8.5\leq log(M/M_{\odot}) \leq 11.1$. The hatched area demarcates the region of surface brightness incompleteness to $log(M/M_{\odot}) < 8.9$ for the SDSS-DR7 galaxy sample used for our analysis. (\textit{Color figure available in the online journal}) \label{Fig2}}
\end{figure*}

\subsection {Spheroid- and disk-dominated galaxies} \label{BDclasses}

\indent Other than the SMF of this global galaxy population, the mass functions of specific sub-groups such as disk- or spheroid-dominated galaxies are also of interest for comparison with simulations. In order to estimate the SMF of these sub-groups, we divide the galaxy population by their bulge-to-total stellar mass fraction, $(B/T)_*$ into four classes, as described below. Galaxy stellar masses derived from SED fitting to multi-band photometry, as in the case of the $MT14$ catalogs, represent a fundamental physical property of the galaxies which may therefore be used for classification, as compared to other schemes based only on the light distribution in a single (e.g., $r$-band) filter. Our adoption of the stellar mass as the fundamental parameter on which to base galaxy classification reflects the trend seen in recent literature where it has been quantitatively tested and well justified, e.g., in \citet{Bluc14, Kelv14a} and \citet{Mous13}. However, even at the outset, we admit that $(B/T)_*$ is a continuously varying parameter in the galaxy population and divisions based on $(B/T)_*$ may seem arbitrary. For example, in a galaxy with $(B/T)_* > 0.8$, classically designated as a spheroid-dominated galaxy, even though the bulk ($\geq 80$\%) of the stellar mass resides in the spheroidal component, the galaxy may still harbor a (minor) disk component. Therefore our artificial segregation of the galaxy sample into four $(B/T)_*$ classes is only for the sake of highlighting any trends in their characteristics.

\indent It must also be pointed out that the galaxy stellar masses we use to compute their $(B/T)_*$ values in this work are taken to be the sum of the stellar masses in their spheroidal, $M_b$ and disk components, $M_d$, which are reported as separate parameters in the $MT14$ catalogs, i.e., $M_g = M_{b+d} = M_b + M_d$. However, the $MT14$ catalogs also contain galaxy stellar masses estimated by fitting them as single entities, $M_g = M_{bd}$. In order to ensure self consistency between these estimates, we compare the two values and find that they match for 545781 galaxies (90.5\%) with differences less than 1$\sigma$ of their combined errors; $MT14$ and \citet{Bluc14} discuss possible reasons for the discrepancy of more than 1$\sigma$ between $M_{b+d}$ and $M_{bd}$ for the other 57341 galaxies. For these galaxies, we take the galaxy stellar mass to be the joint value, $M_{bd}$, and then compute the stellar masses of the individual components using the \textit{photometric} $(B/T)$ value as given in the $i$-band GIM2D catalog; for these galaxies alone we use $(B/T)$ values by their light distribution, and not by their stellar masses. Hence, for these small fraction of galaxies ($<$10\%), $M_b = M_{bd} \times (B/T)_i$ and $M_d = M_{bd} \times [1 - (B/T)_i]$. This consistency check and required correction to the stellar masses of the subset of galaxies is applied before the classification of the galaxies into sub-groups. 

\indent For the classification, we adopt the prescription of \citet{Bluc14}, with one difference, as mentioned below. \citet{Bluc14} base their classification on results from the systematic tests carried out by $MT14$ using 25,000 simulated galaxies spanning the full range of r-band magnitude, half light radius and $(B/T)_*$ values found in the SDSS-DR7 spectroscopic sample (see Appendix B, $MT14$). Using this sample, they not only test the accuracy and precision of the GIM2D spheroid-disk decomposition, but also identify any bias in the recovery of the input parameters. In summary, they find that the threshold for the GIM2D pure S\'{e}rsic parameter, $P_{pS} \leq 0.32$ set by $SL11$ robustly identifies two-component composite galaxies, either spheroid- or disk-dominated. However, in the $P_{pS} > 0.32$ pure S\'{e}rsic sample, for the spheroid-dominated galaxies with high $(B/T)_* > 0.5$, GIM2D sometimes falsely identifies a disk even when the stellar mass contribution of this component is minimal or nil. This issue arises because of the fixed n=4 profile used by GIM2D to fit the spheroids. Where the light profile is steeper than a deVaucouleur's profile \citep{deVau48}, GIM2D compensates by fitting a false exponential disk component. To avoid this misclassification, \citet{Bluc14} define the probability of a false disk, $P_{FD}$, which is based on a combination of the galaxy's axis ratio, $b/a$, the difference in the $(g-r)$ color between the bulge and disk, $\Delta(g-r)_{b,d}$, and the S\'{e}rsic index, $n_s$; all these parameters are taken from the GIM2D catalogs. The overall false-disk probability, $P_{FD}$ is the convolution of these three contributions. For a complete discussion of the rationale behind this, the definition of $P_{FD}$ and the relative weighting of the three contributing parameters, please refer to \citet{Bluc14}, Appendices A-D.

\indent Based on these findings, since the threshold value, $P_{pS} \leq 0.32$ effectively identifies genuine two-component galaxies, we only apply the false-disk correction to galaxies with $P_{pS}$ values higher than this threshold; we differ in this respect from \citet{Bluc14} who apply it to $all$ galaxies, even though they find that the correction is effectively done only to the high $P_{pS}$ population. Within the $P_{pS} > 0.32$ sample, we only select the galaxies with $(B/T)_* > 0.5$ and check their $P_{FD}$ values - following the findings of \citet{Bluc14}, all galaxies with $P_{FD} > 0.2$ are set to be pure spheroids, with $(B/T)_*$=1. A few tens of objects have S\'{e}rsic indices $\gg$4 to which the GIM2D decomposition attempts to fit a fixed n=4 spheroid; to compensate, a false disk is fit in such cases. For this subset of galaxies, we set the galaxy mass to be that obtained using a single S\'{e}rsic fit. For these galaxies, with $(B/T)_*$=1, $M_b = M_g$. For galaxies with $P_{FD} \leq 0.2$, the $(B/T)_*$ values and the stellar masses are left unaltered. 

\begin{figure*} 
\includegraphics[scale=0.7,angle=-90,origin=c]{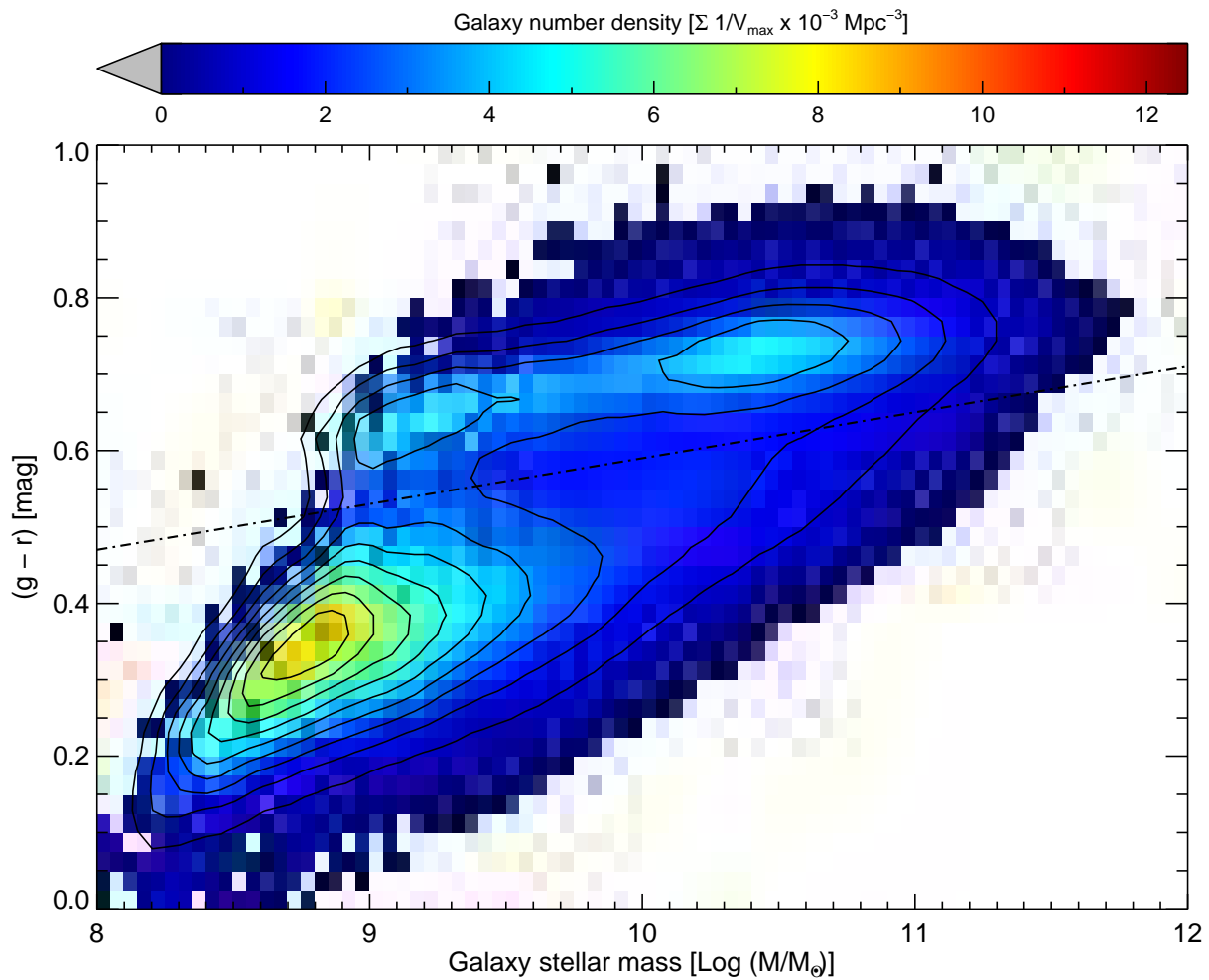}
\caption{Correlation between the galaxy stellar mass and ($g-r$) color of the final sample of galaxies shown as a 2-D density distribution weighted by 1/$V_{max}$. Ten equally spaced contours in the density range 0 to 12.5$\times 10^3 Mpc^{-3}$ are shown overplotted. Regions of enhanced density tracing the \textit{red sequence} of early type galaxies, and the late type galaxies in the \textit{blue cloud} are clearly evident. Also overplotted is the relation, $(g-r)\,=\,0.06\,log(M/M_{\odot})-0.01$ used by \citet{Bluc14} to demarcate the \textit{passive} galaxies from the \textit{active} star forming sub-population. (\textit{Color figure available in the online journal}) 
\label{Fig3}}
\end{figure*}

\indent Having corrected the stellar masses and $(B/T)_*$ values for the presence of false disks, we split the population into four classes from low to high $(B/T)_*$ values as, (i) $(B/T)_*<$ 0.2 (\textit{BT20}), (ii) 0.2$\leq (B/T)_* <$ 0.5 (\textit{BT50}), (iii) 0.5 $\leq (B/T)_* <$ 0.8 (\textit{BT80}), and (iv) $(B/T)_*\geq$ 0.8 (\textit{BT100}). The names used as shorthand to refer to these classes in the following text is shown in italics in parentheses; each name reflects the upper B/T limit (as a percentage) of the corresponding class. We have adopted these classification thresholds because the $(B/T)_*$ division at 0.5 splits the galaxy sample naturally into disk- and spheroid-dominated populations. In each sub-group we further identify the galaxies in which the contribution of the less-dominant component is negligible, e.g., the \textit{BT20} class is disk dominated with only negligible stellar mass in the spheroid; similarly, the \textit{BT100} class has negligible disk contribution. 

For ease of reference, in Table \ref{Tbl1} we list the four classes of galaxies, the selection limits on the stellar mass $(B/T)_*$ ratios we use for each, as well as the number of galaxies in each class. In closing, it must be reiterated that the stellar mass $(B/T)_*$ values have a continuous distribution within the galaxy population. Therefore, our classification and nomenclature are only for the ease of identifying and plotting trends within the population, which may then be interpolated for the corresponding properties of galaxies with intermediate values of $(B/T)_*$.    

\subsection {Characteristics of the galaxy, spheroid and disk catalogs} \label{CatProps}

In order to understand the general characteristics of our galaxy sample as well as to test for any systematic bias in our estimates of the SMF and associated analyses, we first present pertinent properties of the population in our catalogs. First, the redshift distribution of the galaxies in the GIM2D catalogs which meet our selection criteria is shown in Figure \ref{Fig1}(\textit{left}); for ease of comparison, the y-axis has been scaled to show the relative number of galaxies in each z=0.01 bin with respect to the total number of galaxies, as well as the actual number of galaxies in each redshift bin. The hatched areas on the plot demarcate our chosen redshift range within which the distribution peaks at a median redshift, z=0.099. Note that the number of galaxies in each redshift bin ($\Delta z=0.01$) is well over a few thousands to tens of thousands of galaxies. This number per redshift bin equals or exceeds what has been the total sample size in several earlier similar studies of galaxy SMF; this impressive sample size provided by SDSS permits us to derive statistically robust inferences of the characteristics of this representative, low-redshift galaxy population. 

\indent A histogram of galaxy stellar masses, in bins of 0.2 dex in solar mass, is shown in Figure \ref{Fig1}(\textit{right}), adopting a similar absolute and relative scaling for the y-axis as used for the redshift plot. The distribution peaks in the bin centered at $10^{10.8}M_{\odot}$, and declines steeply toward $10^{12}M_{\odot}$ in keeping with the drop in the number of massive galaxies in this mass range in the local universe. The decline in the relative distribution at lower stellar masses is due to the decrease in the completeness of the survey due to the drop in the surface brightness of these low mass galaxies. The effect of this incompleteness is evident in the decline in the galaxy SMF as discussed under \S \ref{bd2gSM}. In \S \ref{mur_comp}, we use the weighted bivariate distribution of the $r$-band surface brightness and the stellar mass of the galaxy sample to determine the stellar mass below which the robustness of our inferences may be affected by the incompleteness of the SDSS galaxy sample. The grey shaded region in Figure \ref{Fig1}(\textit{right}) demarcates this region of incompleteness. This precautionary shaded region indicating sample incompleteness is shown in all subsequent plots.  

\begin{figure*} 
\includegraphics[scale=0.7]{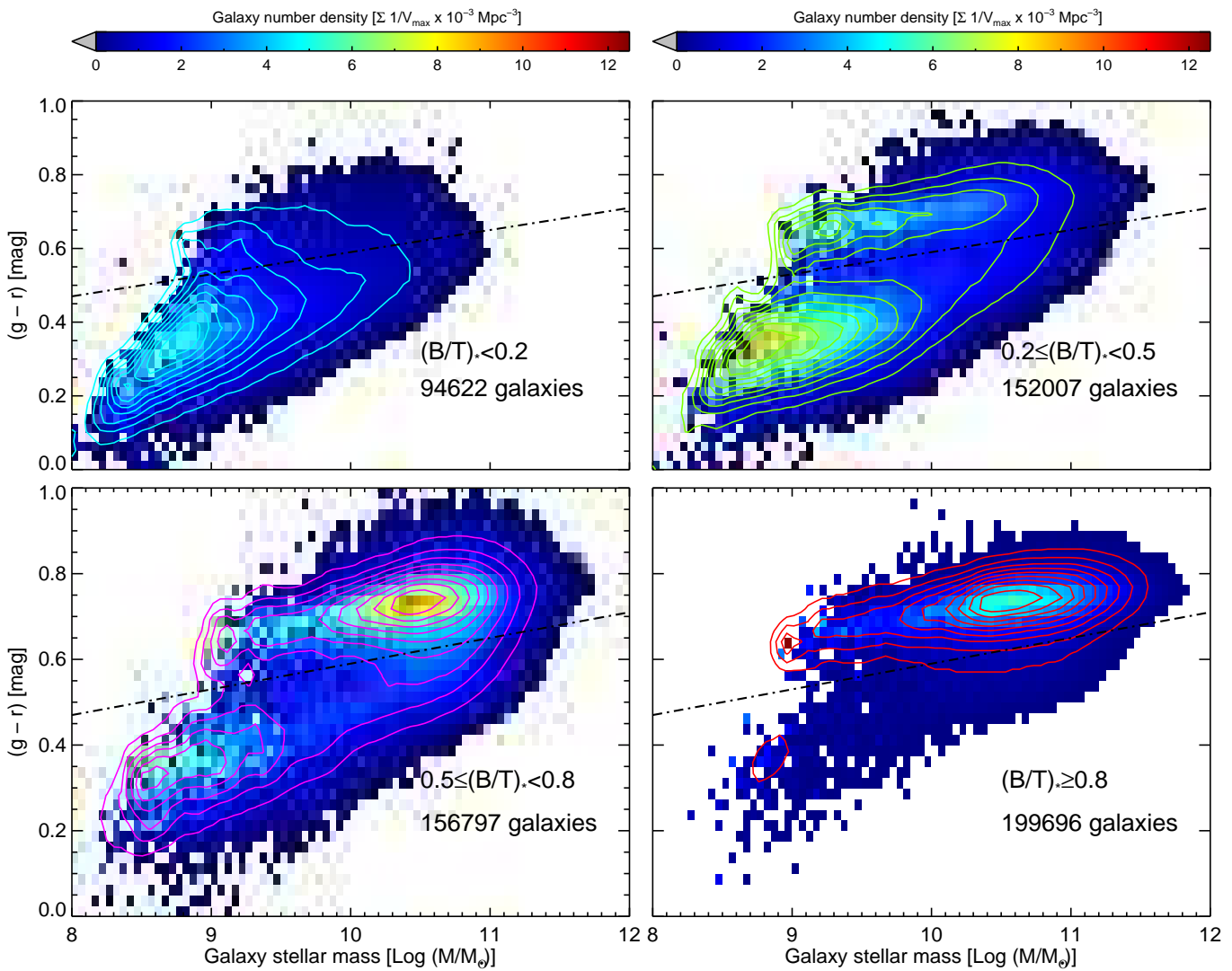}
\caption{Panels showing the stellar mass to ($g-r$) colour distributions for the four $(B/T)_*$ classes in order to highlight where these populations lie with respect to the \emph{red sequence} and the \emph{blue cloud}. Overplotted are contours tracing the respective peak values in each plot to help highlight the regions of higher density. Also overplotted on each panel is the $(g-r)$ to stellar mass relation (given in Figure \ref{Fig3}) demarcating the \textit{passive} and \textit{active} sub-populations. The migration of the high density region from bluer, low stellar mass regions in the disk-dominant classes through the \textit{green valley} to the red, high mass ridge defining the \emph{red sequence} of the spheroids is evident. (\textit{Color figure available in the online journal}) \label{Fig4}}
\end{figure*}

\subsection{SDSS-DR7 surface brightness completeness}\label{mur_comp}

Quantifying the completeness limits of the SDSS-DR7 survey is critical to establish the stellar mass range over which our estimated SMFs and associated results apply. Regarding the importance of estimating the completeness of any survey, \citet{Cros01} used a sample of 45000 galaxies from the Two-Degree Field Galaxy Redshift Survey (2dFGRS) to show that surface brightness completeness of a survey, and to a lesser extent cosmic variance, have significant impact on the estimated galaxy luminosity function and luminosity density. Incorporating a correction for these effects leads to a 37\% increase in the estimated luminosity density, of which 35\% is from the surface brightness correction and the remaining by accounting for galaxy clustering. \citet{Cros02} find that convolving the Schechter function \citep{Sche76} with a gaussian describing the \textit{bivariate brightness distribution} (BBD) of the sample leads to better consistency between the derived Schechter parameters and luminosity density from various surveys.

\indent \citet{Bald08} have highlighted that estimates of the galaxy SMF and stellar mass density are also affected by the surface brightness incompleteness of any survey. Based on the SDSS NYU-VAGC catalog (New York University-Value Added Galaxy Catalog) \citep{Blan05b}, they conclude that the decline in the SMF for stellar masses $< 10^9 M_{\odot}$ is due to surface brightness incompleteness as well as the presence of large scale structure. Earlier estimates by \citet{Blan05a} put the surface brightness completeness of the SDSS-DR2 at $\sim$70\% for $\mu_r \leq $23 mag.arcsec$^{-2}$, the limiting surface brightness for the SDSS spectroscopic galaxy sample, which we use here for the SMF estimation. 

\indent Therefore, in order to determine the limiting stellar mass at which the surface brightness incompleteness in the SDSS-DR7 becomes significant, we too adopt the \citet{Bald08} approach based on the correlation between stellar mass and $r$-band surface brightness, $\mu_r$ (see their Figure 4). In Figure \ref{Fig2} we plot the $1/V{max}$ weighted \textit{bivariate brightness distribution} (BBD, \citet{Cros01}) between galaxy stellar mass and $\mu_r$ taken from the SDSS database. For the surface brightness distribution in each stellar mass bin, we then fit a gaussian and overplot the estimated central value and 1$\sigma$ uncertainty for the stellar mass range $8.5\leq log(M/M_{\odot}) \leq 11.1$. In the stellar mass range, $8.9\leq log(M/M_{\odot}) \leq 10.9$, the relation is seen to be linear with a slope of -0.94 and reduced $\chi^2_\nu \ll 1$, as shown by the dashed line. However, at lower stellar masses, $log(M/M_{\odot})<8.9$ there is a clear departure of the central value from the linear fit with a rapid increase in $\chi^2_\nu \geq 5$. Therefore, we set $log(M/M_{\odot})=8.9$ to be the lower end of the stellar mass range from which we are able to draw robust inferences. We address the effect of large scale structure on the SMF and any residual effects even after the $1/V_{max}$ correction in Appendix \ref{LSS}.

\subsection{1/$V_{max}$ completeness correction}\label{VmaxCorr}

\indent We use the non-parametric 1/$V_{max}$ approach \citep{Schm68, Felt76, Eale93} for completeness correction to account for the survey depth. The interested reader is referred to the review by \citet{John11} for a detailed discussion regarding the applicability of this method to compute luminosity functions and SMF estimates. The $V_{max}$ values were computed using the method described in \citet{Sima11} (\S 3.6, Eqn. 7) but for the redshift limits $z_{min}=0.02$ and $z_{max}$=0.2 adopted in our study. In computing Vmax, appropriate $k$-corrections using the \citet{Blan07} approach have been applied. However, given the small redshift span ($\Delta z< 0.2$) of our low redshift galaxy sample, no correction for luminosity evolution was applied. If the galaxies are uniformly distributed over the survey volume, the assumption under which the $1/V_{max}$ completeness correction works, the median value of $V/V_{max}$ should be 0.5, where $V$ is the SDSS spectroscopic survey volume measured up to the redshift of each galaxy. For our galaxy sample, the median value of $V/V_{max} = 0.497$, thus justifying the use of the $1/V_{max}$ approach for completeness correction.
  
\indent Along with this $1/V_{max}$ completeness correction, we also investigated a correction for incompleteness due to \textit{fiber collisions} in the SDSS-DR7 spectroscopic sample. In the sample selection for the SDSS spectroscopic survey, galaxies at spatial separations less than 55$\asec$ could not be co-observed due to the finite diameter of the cladding on the optic fibers feeding the spectrographs \citep{Stra02}. Due to this limitation imposed by \textit{fiber collisions}, \citet{Stra02} estimated that $\sim$6\% of the galaxy sample from the photometric survey, which pass the magnitude and other selection criteria for spectroscopic follow-up, could still not be observed. In order to estimate the consequent level of incompleteness in the SDSS-DR7 spectroscopic catalog, we compared the number of galaxies in the photometric catalog which meet the spectroscopic selection criteria but are not found in the DR7 spectroscopic catalog. Based on this, we estimate that the incompleteness due to fiber collisions stands at $<$4\% in the SDSS-DR7. This improvement in completeness is because galaxies which were missed in earlier runs have subsequently been observed with SDSS plates for other spectroscopic surveys. 

\indent To correct for this \textit{target selection rate} (TSR) we adopted the method used by \citet{Li09}, and \citet{Peng10} using the spectroscopic completeness parameter obtained from the New York University - Value Added Galaxy Catalog (NYU-VAGC)\footnote{http://sdss.physics.nyu.edu/vagc/} \citep{Blan05b}. As defined in those earlier investigations, the combined completeness correction weighting is 1/$V_{max} \times$ 1/TSR. We then compared the overall shape and magnitudes of the resulting galaxy SMF with and without the TSR correction, as well as the fitted values of the Schechter parameters, and noted only  minor differences. The change in the stellar mass density was $<$2\%. Regarding the TSR correction parameter, since no spectroscopic information is available for the galaxies that were not observed due to fiber collision, the TSR correction assumes that the missed galaxy has the same redshift and other spectral properties as the closest neighbour which was observed. In the high density environments where fiber collisions are most likely to occur, this assumption may preferentially bias stellar masses toward higher values since the more massive, hence brighter galaxy would have been selected for spectroscopic follow-up instead of the nearby companion. In order to avoid such a bias, and with the aim to keep the number of corrections applied to our sample to a minimum, we have decided to leave out the TSR correction in the results quoted in this paper. An increase of 2$\%$ due to TSR correction may be applied to the stellar mass densities we have quoted, should an interested reader wish to include this correction.
 
\begin{figure*} 
\includegraphics[scale=0.75]{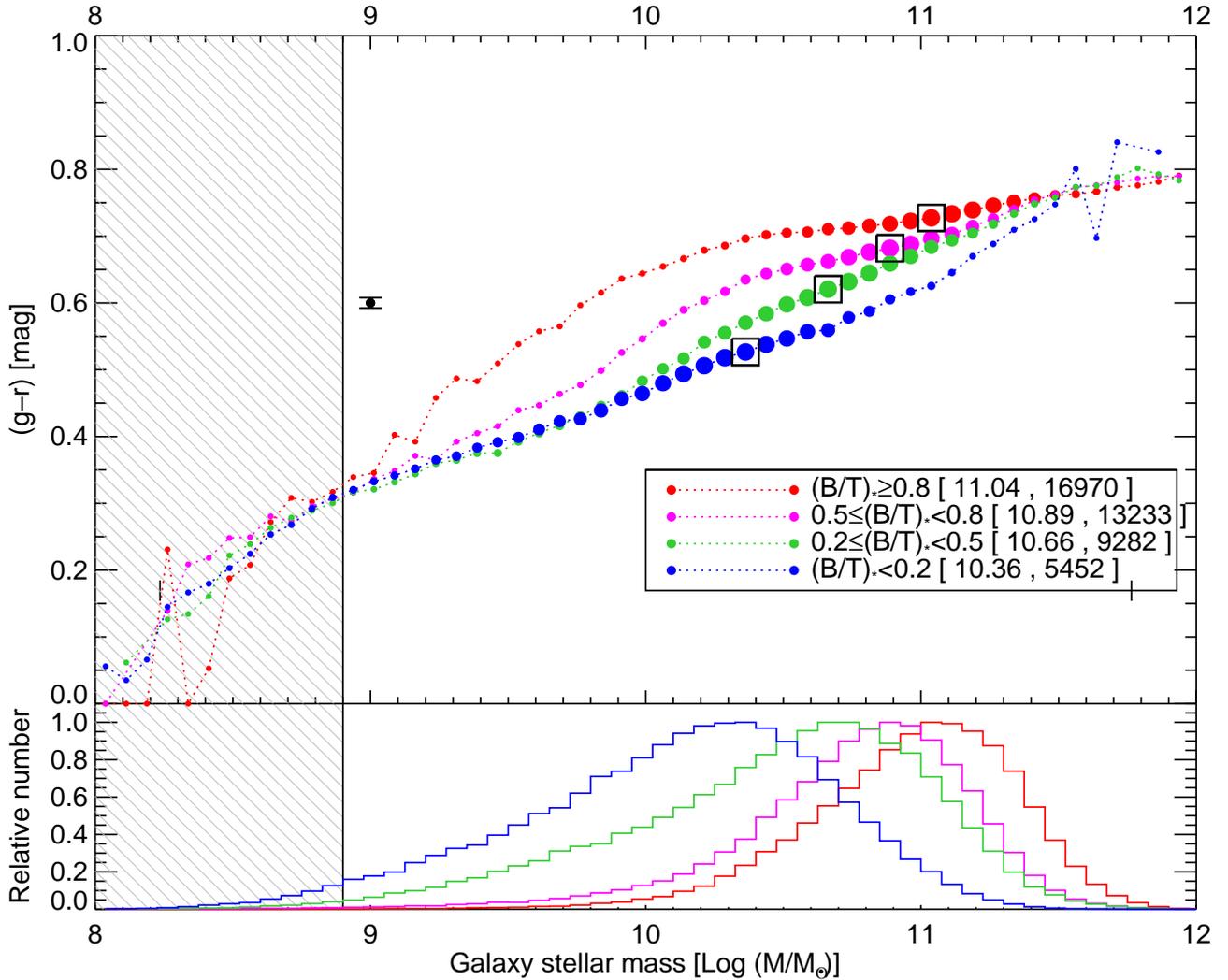}
\caption{(\textit{Upper}) The migration of galaxies' colours from the \textit{blue cloud} to the \textit{red sequence} shown as a function of galaxy stellar mass for the four sub-groups of galaxies defined in \S \ref{BDclasses}. The symbol sizes in each plot are scaled with respect to the maximum number of galaxies per bin for that class (highlighted with a square symbol). The maximum number per bin for each class is shown in the legend along with the stellar mass bin in which it occurs. In order to avoid crowding on the plots, a single representative error symbol is used to indicate the measured scatter in the ($g-r$) colour in any stellar mass bin in the range of interest. (\textit{Lower}) The corresponding histograms normalized with respect to the maximum number per bin for each class to visually highlight the stellar mass at which it occurs. (\textit{Color figure available in the online journal}) \label{Fig5}}
\end{figure*}

\subsection {Color migration from disks to spheroids} \label{ClrMig}

\indent Using the 1/$V_{max}$ correction, we explore the color distribution of our galaxy sample as a function of stellar mass and $(B/T)_*$ ratio. Shown in Figure \ref{Fig3} is the 1/$V_{max}$ weighted density distribution of the galaxy stellar mass, $M$ against the corresponding ($g-r$) color of our full sample of galaxies. The overplotted contours help highlight the regions of higher density, as well as the directions of the density gradients. This completeness corrected density distribution traces the early type galaxies along the \emph{red sequence} and the late type population in the \emph{blue cloud} as regions of enhanced density. Overplotted in Figure \ref{Fig3} is the \citet{Bluc14} correlation between the galaxy color and its stellar mass which demarcates the \textit{passive} galaxies from the \textit{active}, star-forming sub-population. The density peak for the redder, dominantly spheroidal population lying on the redward side of this line, extends toward higher stellar masses, while the \emph{blue cloud} galaxies are preferentially lower mass systems. However, the underlying density distribution shows that the number of  \emph{blue cloud} galaxies per 2D bin is twice higher than those along the ridge of the  \emph{red sequence}, in line with the hierarchical galaxy evolution scenario in which the less luminous blue galaxies merge to build the redder, passive systems which populate the \emph{red sequence}.  

\indent Other than the well known bimodality in colors within the galaxy population, using the large, homogeneous SDSS population we aim to trace the expected migration of galaxies through the \textit{green valley} \citep{Stra01} between the two peaks represented by the \emph{blue cloud} and the \emph{red sequence}. In Figure \ref{Fig4}, we identify regions in the stellar mass versus color plane populated by the four structural subgroups defined in \S \ref{BDclasses} and present them in a format similar to the one used for the full population, Figure \ref{Fig3}. Overplotted on each correlation are contours to help highlight regions of higher density; the underlying distribution was smoothed with a 3x3 Gaussian filter prior to contouring. On each panel, the line demarcating the passive and active star-forming galaxies has also been overplotted for reference.

\indent Tracing the location of the higher density (inner) contour in each sub-group beginning with the  \textit{BT20} class in the top, left panel, it is seen that the high density region moves systematically to redder colours and higher stellar masses, following the evolution of the bulk of the galaxies making up that class. The segregation of the high \textit{BT100} class from the other classes is distinct, with essentially all these red galaxies ($\geq 95$\%) lying along the \emph{red sequence}, while an equivalent fraction of the \textit{BT20} class galaxies are restricted to the \emph{blue cloud}. However, in the case of the composites, there is a generous overlap between the overall distributions of the two intermediate classes of galaxies, but the migration of the peak in the inner contour toward the \emph{red sequence} with increasing $(B/T)_*$ is distinctly visible. The region of overlap between these two sub-groups shows the transition occurring in the \textit{green valley} \citep{Stra01} due to merging and secular processes in these two-component systems driving the migration toward the high S\'{e}rsic index objects in the high \textit{BT100} class lying along the sharp ridge of the \emph{red sequence}. This is consistent with the well known \textit{color-morphology} (essentially \textit{color-density}) relation \citep{Dres80, Post84, Whit91, Pele96} that all passive (red) galaxies have a significant spheroidal component, as confirmed by \citet{Bluc14, Alpa15, Bren15, Bell12} and \citet{Bell08} amongst others.

\begin{figure*} 
\includegraphics[scale=0.75]{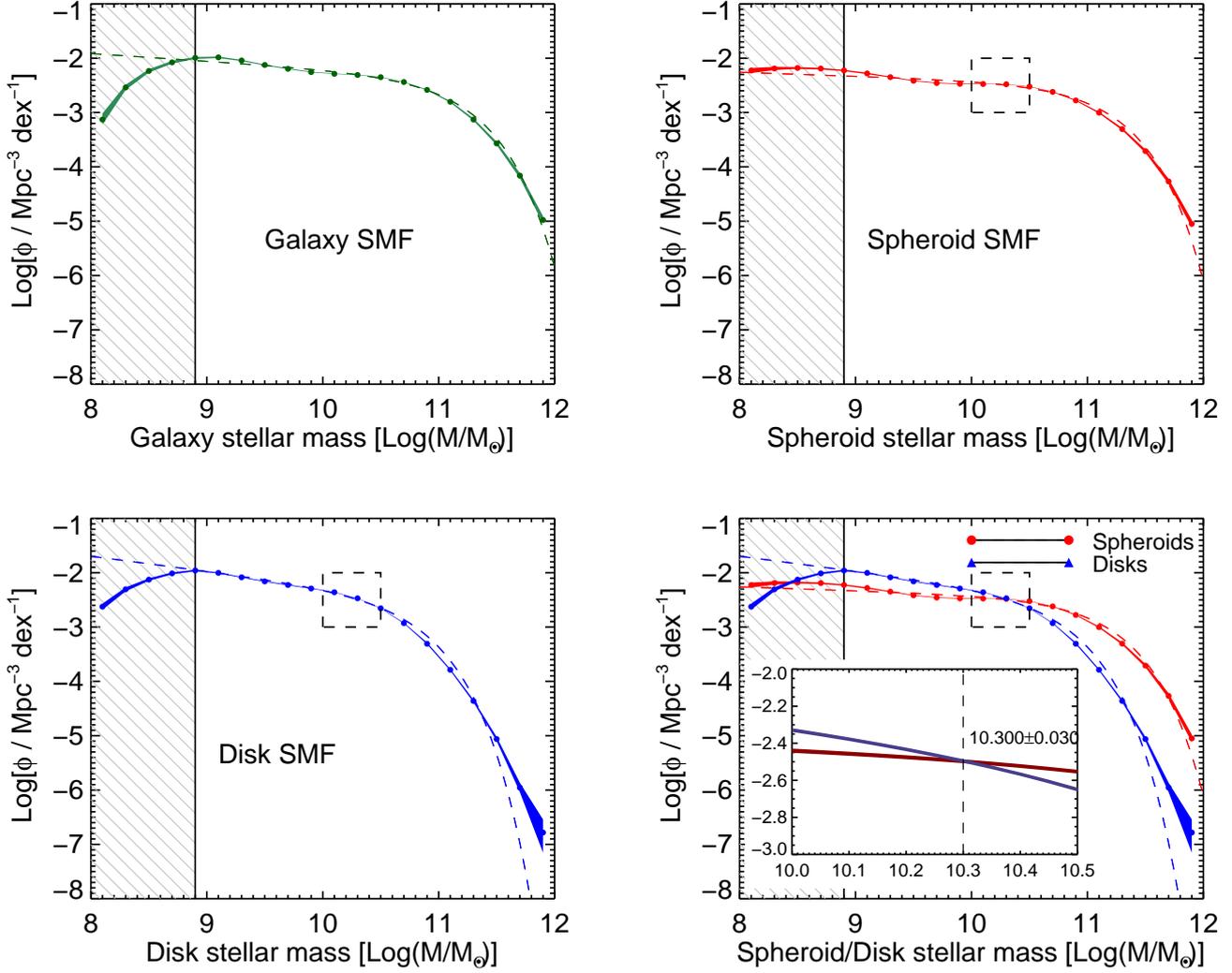}
\caption{(\textit{Top, left}) The SMF of the 603,122 galaxies in our SDSS-DR7 sample which meet all our selection criteria. The dashed line represents the Schechter function fit to the SMF, with the values of the Schechter parameters given in Table \ref{Tbl2a}. Due to incompleteness at the low mass end, the hatched area has been excluded from the fit (see \S\ref{bd2gSM}). In this and all subsequent SMF plots, the width of each trace represents the corresponding total uncertainty. (\textit{Top, right}) The SMF representing the stellar mass in the spheroidal components of all the galaxies in our sample. The dashed region demarcates the inset in the bottom, right panel for the \textit{crossover stellar mass}. (\textit{Bottom, left}) The SMF representing the stellar mass in the disk components of the galaxies in our sample.  (\textit{Bottom, right}) The SMFs of the spheroidal components and the disk components overplotted to indicate the \textit{crossover stellar mass}, which is annotated in the zoomed inset plot. (\textit{Color figure available in the online journal}) \label{Fig6}}
\end{figure*}

\indent In order to further explore this colour transition as a function of the galaxy stellar mass and the $(B/T)_*$ stellar mass ratio, in the upper panel of Figure \ref{Fig5} we trace the color distribution of the galaxies in each of these four classes binned by their stellar mass in 0.075 log($M/M_{\odot}$) bins. In order to identify the high density regions in these traces, the plot symbols are scaled relative to the maximum number of galaxies per bin for that class; hence the bin with the maximum number of galaxies for each class is indicated with the largest symbol, which is also annotated with a square as a visual highlight. In order to avoid crowding in the plot, a single representative point with error bars is indicated by the black (arbitrary) point. The corresponding histograms are plotted in the lower panel, each normalized with respect to the maximum number per bin for that class. The histograms highlight the location of the maximum and also the distribution of galaxies about that maximum for each class. 

\indent The ($g-r$) colour of disk-dominated galaxies may be affected by internal dust reddening when viewed at low inclination angles. In order to test for such a bias, we regenerate the trace of the stellar mass to ($g-r$) colour correlation for the disk-dominated galaxies, selecting only those with disk inclinations $\geq30$ degrees from being edge-on. The trends remain consistent with those for the full sample. Therefore, in Figure \ref{Fig5} we plot the correlations for the full sample.   

 \begin{table*}
 \begin{center}
 \caption{Schechter parameters for SMF of galaxies, spheroids and disks.} \label{Tbl2a}
 \scriptsize
 {\renewcommand{\arraystretch}{2.0}
 \begin{tabular}{|L{2cm}|C{3cm}|C{2cm}|C{2cm}|C{2cm}|C{2cm}|}
\hline
 \hline
 Population & $\phi^*$ & $log_{10}\;(M^*/M_{\odot})$ & $\alpha$ & $\rho_{Sch}$ & $\mathbf{\rho_{SMF}}$ \\
  \hline
  & [$10^{-3}\; Mpc^{-3}\; dex^{-1}$] & [ - ] & [ - ] & [$10^8\;M_{\odot}\;Mpc^{-3}$] & [$10^8\;M_{\odot}\;Mpc^{-3}$] \\
 \hline
 Galaxies & $1.911^{+0.045}_{-0.045}$ & $11.116^{+0.011}_{-0.011}$ & $-1.145^{+0.008}_{-0.008}$ & $2.760^{+0.109}_{-0.109}$ &  $2.670^{+0.109}_{-0.110}$ \\ 
\hline
Spheroidal components & $1.424^{+0.027}_{-0.028}$ & $11.085^{+0.010}_{-0.010}$ & $-1.073^{+0.007}_{-0.006}$ & $1.812^{+0.062}_{-0.062}$ & $1.687^{+0.063}_{-0.062}$ \\ 
\hline
Disk components & $1.582^{+0.029}_{-0.030}$ & $10.707^{+0.006}_{-0.006}$ & $-1.277^{+0.007}_{-0.007}$ & $1.006^{+0.029}_{-0.029}$ &  $0.910^{+0.029}_{-0.029}$ \\
\hline
\hline
\end{tabular}
}
\end{center}
\begin{flushleft}
\footnotesize{Fitted Schechter parameters describing the SMF of the low redshift ($z\sim0.1$) population of galaxies from the SDSS-DR7, and their spheroidal and disk components based on the $SL11$ and $MT14$ catalogs. The Schechter parameter values are obtained using a standard maximum likelihood fit to the SMFs, described in \S \ref{SMFbd}. The quoted errors are $16^{th}$ and $84^{th}$ quartile values obtained by MC sampling. The last two columns list the corresponding stellar mass densities in the local univers, with $\rho_{Sch}$ being the integrated Schechter function, and $\rho_{SMF}$ obtained by numerical integration of the SMF. }
\end{flushleft}
 \end{table*}
 \normalsize 

\indent At the low end of the stellar mass distribution, with log$(M/M_{\odot}) \sim$ 9, galaxies in all four classes show the same mean ($g-r$) color of $\sim0.3$, with the width of the scatter being equal as well. There is a general trend toward bluer colors with decreasing stellar mass in all four classes, but the increasing incompleteness in this mass range does not permit a firm inference. On the other hand, with increasing stellar mass, the galaxies in the highest $(B/T)_*$ class begin to exhibit a clear redward trend at log$(M/M_{\odot}) \geq$ 9, and are well established on the \emph{red sequence} for $M\geq 10^{10}M_{\odot}$ with mean $(g-r)\geq0.64$ mag.  For the low $(B/T)_*$ class, the migration to redder colors is gradual and monotonic, with galaxies of higher stellar masses being redder than the lower mass galaxies in that class, indicative of an increasing fraction of more evolved stellar populations contributing to their stellar mass budget. For the composite galaxies, the redward trend occurs at stellar masses log$(M/M_{\odot}) \geq$ 9.3 for the \textit{BT80} class, while the migration for their \textit{BT50} counterparts occurs only about $\sim0.8$ dex later. It is worth noting that even while on the red sequence, the colors of the galaxies consistently show a hierarchical segregation, with the higher B/T galaxies being $\sim0.1$mag $redder$ than their disk-dominated counterparts, indicative of an older, evolved stellar population versus perhaps residual star formation in the lower B/T galaxies. This hierarchy decreases with increasing stellar masses so that at the highest mass bins used in this comparison, the difference in colors, $\Delta(g-r)$ between the four classes is less than the corresponding 1$\sigma$ uncertainties. 

\section{Results}\label{Results}
\subsection {Galaxy stellar mass function} \label{SMFg}

We obtain the SMF of this representative sample of z$\sim$0.1 galaxies using the standard, non-parametric $1/V_{max}$ approach \citep{Eale93, Felt76, Schm68}, with the SMF given by,
\begin{equation}
\label{VmaxEqn}
\phi(\mathrm{log(M)})\,d(\mathrm{log(M)}) = \sum_{i=1}^{n}\,\frac{{\mathrm{w}_i}}{V_{max,i}}
\end{equation}
The $V_{max}$ values represent the maximum survey volume over which a given galaxy could lie and still satisfy the magnitude limits of the survey. For the SMF, the $V_{max}$ values have been computed for the redshift range we have adopted, $0.02 \leq z \leq 0.2$, as described in \S \ref{CatProps}. Appropriate $k$-corrections following the \citet{Blan07} methodology have been used in the $V_{max}$ computation; however, given the small redshift range, no evolutionary corrections have been applied. The weighting functions, $w_i$ in Equation~(\ref{VmaxEqn}) may be used to incorporate additional corrections such as the \textit{target selection rate} (TSR); however, for reasons discussed in \S \ref{CatProps}, we have not included TSR corrections for the SMF estimates given here, hence we set $w_i = 1$ for the galaxies in our sample. 

\indent The galaxy SMF estimated using the SDSS-DR7 galaxy sample at median redshift, $z\sim0.1$ is plotted in Figure \ref{Fig6} (\textit{upper left}). The uncertainties in the stellar masses in the $MT14$ catalogs are propagated to the SMF by applying Monte-Carlo (MC) sampling using the two-sided stellar mass uncertainties given for each galaxy. The MC sampling is repeated 100 times, and the $16^{th}$ and $84^{th}$ quartile values in each stellar mass bin are used as the corresponding uncertainties in the SMF. Given the sample size with a few $10^4$ galaxies per stellar mass bin, the contribution from the statistical (Poisson) uncertainty is small. On the other hand, systematic uncertainties arising from assumptions made by $MT14$ in the flexible stellar population synthesis \citep{Conr10a} are the dominant contributors to the error budget, as described in \S\ref{SMCats}. Therefore, in the appendices we estimate the contributions of the principal factors to the overall systematic uncertainties in the SMF, with Appendix \ref{IntcompSMFg} focussed on systematics due to selectively combining four different galaxy stellar masses from $MT14$ to correct for various effects described in \S \ref{BDclasses}. Appendix \ref{SystErrChksSMFg} is related to the assumptions made in the 11 main parameters governing the SED fits for the stellar mass estimates in $MT14$. The overall uncertainty in our SMF estimates is taken to be the contributions from the MC sampling, the Poisson uncertainty and these two systematic uncertainties summed in quadrature. The filled region around the SMF in Figure \ref{Fig6} indicates the width of the uncertainty contour as a function of stellar mass.
   
\indent We then fit the corresponding Schechter function \citep{Sche76} to the SMF weighted with the combined uncertainties. We use the single Schechter function given by,
\begin{multline}
\label{SchFnEqn}
\phi(M)dM =  ln(10) \phi^* \Big \{ 10^{(M-M^*)(\alpha+1)} \Big \} \\
  exp\Big (-10^{(M-M^*)} \Big ) dM
\end{multline}

where for brevity we have defined, $M = \mathrm{log}(M/M_{\odot})$. We use the standard Levenberg-Marquardt minimization technique \citep{Pres92} to obtain the normalization constant, $\phi^*$, the faint end slope, $\alpha$ and the characteristic stellar mass, $M^*$. Due to the increasing incompleteness in the SDSS survey at the lower end of the stellar mass range, log$(M/M_\odot) \leq$8.9, as discussed in \S \ref{mur_comp}, the galaxy SMF turns down and the corresponding width of the uncertainty region increases. During the Schechter function fit, there is a corresponding significant increase in the reduced $\chi_\nu^2$ value when we include these stellar mass bins in the fits. Therefore, in order to avoid misleading trends in the Schechter fits and our inferences therefrom due to these higher uncertainties, we do not include the region log$(M/M_\odot)\leq$8.9 in the fits (indicated by the hatched region in Figure \ref{Fig6} and in all subsequent plots). 

\indent During the fitting procedure, we take into account all the systematic and statistical uncertainties in the SMF mentioned above; the stellar mass however, being the independent variable, is left unweighted. To account for the systematic uncertainties in the stellar masses, we fit the Schechter function to the 100 MC realizations of the SMF along with the associated total uncertainties to obtain the corresponding sampling of the Schechter parameters. The median and interquartile distances of the fitted Schechter parameters from the 100 SMF realizations are listed in Table \ref{Tbl2a} as the values of these parameters and their corresponding uncertainties. The fitted Schechter function is overplotted on the galaxy SMF in Figure \ref{Fig6} (\textit{top, left panel}) 

\indent Also listed in the last two columns of Table \ref{Tbl2a} are the stellar mass densities and associated uncertainties for the galaxies. The column $\rho_{SMF}$ is obtained by numerically integrating the SMF over the stellar mass range, 8.9 $\leq$log$(M/M_\odot) \leq$12, where the galaxy sample is unaffected by surface brightness incompleteness, described in \S \ref{mur_comp}. Related to the findings of \citet{Bern13} discussed in \S \ref{ExtcompSMFg} below, this also accounts for the differences between the SMF and corresponding Schechter fit \citep{Sche76} in the higher stellar mass bins. 

\indent In order to account for the contribution from the low end of the stellar mass range to the stellar mass density, we extrapolate the fitted Schechter function to the region of incompleteness, and integrate over the range, 8 $\leq$log$(M/M_\odot) \leq$12 to obtain $\rho_{Sch}$. We use the quartile values from 100 MC realizations of the SMF to provide the associated uncertainties in these stellar mass densities. The differences between these two mass densities (taking into account the different mass ranges over which they apply) is $\sim$3\% for galaxies. By comparing the values of $\rho_{Sch}$ over the two stellar mass ranges, the contribution of the low mass galaxies, which lie in the region of incompleteness, log$(M/M_\odot)\leq$ 8.9 is found to be 1.23\%.  

\indent With a characteristic mass $M^* = 11.109^{\textsc{\tiny +0.010}}_{\textsc{\tiny -0.010}}$ and faint end slope, $\alpha = -1.142^{\textsc{\tiny +0.008}}_{\textsc{\tiny -0.007}}$, our estimate of the galaxy SMF is consistent with other recent estimates in literature within the combined total (= statistical + systematic) uncertainties of both measurements, as discussed below. The corresponding stellar mass density of galaxies, $\rho_{Sch}$ in the local universe is $2.760^{\textsc{\tiny +0.109}}_{\textsc{\tiny -0.109}}$ in units of $10^8 M_{\odot} Mpc^{-3}$ in the stellar mass range, 8 $\leq$log$(M/M_{\odot}) \leq$12. Restricting this to the estimate where the SDSS sample is complete, the mass density, $\rho_{SMF}$ = $(2.670^{\textsc{\tiny +0.109}}_{\textsc{\tiny -0.110}})$ $\times 10^8 M_{\odot}$. We also compare these density estimates with representative results from recent literature.  

\subsection{External consistency checks: Galaxy SMF} \label{ExtcompSMFg}

\begin{figure*} 
\includegraphics[scale=0.75]{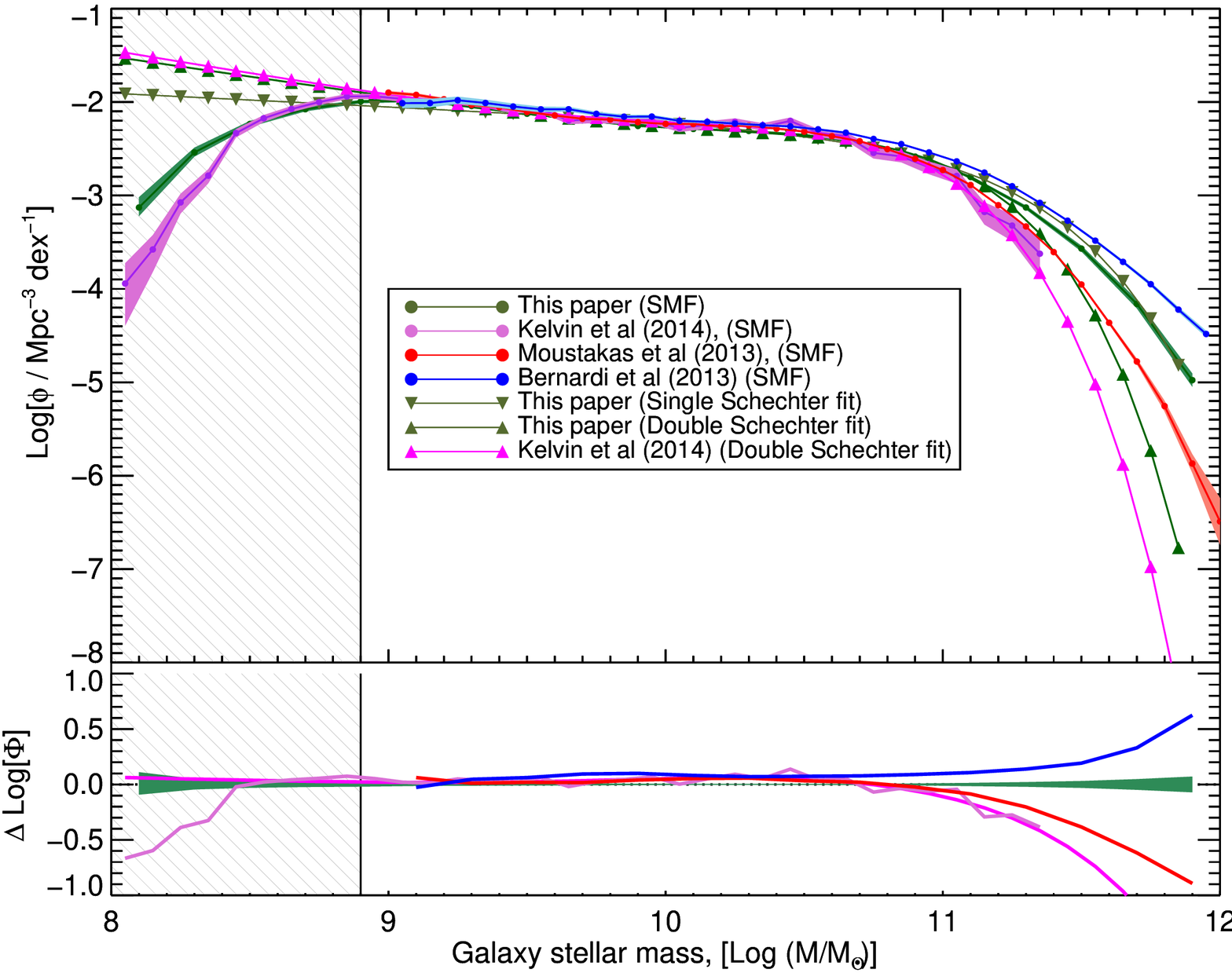}
 \caption{(\textit{Upper panel}) The SMF and fitted Schechter functions (single and double) for our galaxy sample compared with corresponding SMF or Schechter function fits obtained for the GAMA survey \citep{Kelv14a}, by \citet{Mous13} for the PRIMUS survey, and by \citet{Bern13} for a SDSS-DR7 sample. The SMF of the GAMA survey uses data provided by \citep{Kelv14a} (\textit{priv. comm}). A double Schechter function is also used for the the GAMA survey comparison. For the PRIMUS survey, the SMF and associated uncertainties listed by \citet{Mous13} (Table 3) are plotted.The \citet{Bern13} SMF is plotted using their online data table for their S\'{e}rsic light profile fit. (\textit{Lower panel}) Difference between the GAMA SMF, the double Schechter function fit, the PRIMUS SMF and the SDSS-DR7 SMF and our corresponding single, double Schechter function fit or SMF. The uncertainty contour in our SMF estimate is indicated by the green shaded region. (\textit{Color figure available in the online journal}) \label{Fig7}}
\end{figure*}

The three recent measurements chosen for this comparison are from the \emph{GAMA} survey \citep{Kelv14a}, the \emph{PRIMUS} survey \citep{Mous13} as well as from a sample drawn from SDSS-DR7 \citep{Bern13, Bern10}. In the publications cited, SMF results from other earlier surveys have been been used to benchmark each of these estimates, hence our comparisons may be further augmented with the discussions given therein. 

\indent \citet{Kelv14a} estimate the galaxy SMF using 2711 galaxies from the GAMA survey first data release \citep{Driv11}. GAMA is an ongoing multi-wavelength photometric and spectroscopic survey overlapping the southern SDSS footprint, and targets galaxies with an extinction corrected SDSS r-band Petrosian magnitude, $r \leq 19.4$ mag and within a redshift range $0.025 < z < 0.06$. The stellar masses are estimated by fitting the optical and near infrared photometry to a suite of synthetic stellar templates generated with the Bruzual \& Charlot model \citep{Bruz03} with a Chabrier IMF \citep{Chab03}. The PRIMUS results \citep{Mous13} pertain to the SMF obtained for a sample of 170,000 SDSS galaxies which lie in the redshift range $0.01 \leq z \leq 0.2$, thus well matched with our median redshift of $z\sim0.1$. Stellar masses are estimated by SED fitting to the available multi-band UV, optical and near-IR photometry.

\indent The SMF of the GAMA survey shown in Figure \ref{Fig7} represents the results presented in \citet{Kelv14a} and was provided by the principal author (\textit{Kelvin,~L.~S., priv. comm.}). For the PRIMUS survey, we plot the SMF and associated uncertainties tabulated in Table 3 of \citet{Mous13} as a function of the stellar mass within the range 9$ \leq log(M/M_\odot) \leq$12. Other than the comparison of the SMFs, we also compare the Schechter fits to these estimates. Recent SMF studies such as \citet{Kelv14a} argue in favour of a double Schechter function being a better fit for the galaxy SMF in order to adequately capture the steeper faint end slope. We do note this marginal rise in our SMF in the three bins prior to the onset of the drop due to incompleteness at stellar masses, log$(M/M_\odot) \leq$8.9. We therefore also plot a double Schechter function fit to our galaxy SMF estimate and thus carry out a fair comparison with \citet{Kelv14a} result. 

\indent For our third comparison set, we use the online data table provided by \citet{Bern13} for the SMF of their SDSS-DR7 sample estimated using stellar masses from S\'{e}rsic fits to the galaxy light profiles. \citet{Bern13} caution about the sensitive dependency of the estimated galaxy stellar masses (and hence the SMF) on the profile assumed to describe the observed light distribution. Further, they mention that additional uncertainty arises also from the assumed mass-to-light ratio used to convert the measured luminosity to stellar mass. They therefore correct the SDSS magnitudes, obtained in the standard pipeline with a Petrosian profile fit, using an improved S\'{e}rsic fit and recover a stellar mass density 20\% higher than their previous estimate with the same galaxy sample \citep{Bern10}. They find the differences to be particularly evident at higher stellar masses, $M\geq 6\times10^{11}M_{\odot}$. Based on these results, they suggest an analytic function which is a better representation of the underlying SMF than the standard Schechter function. In a recent publication, \citet{DSou15} apply a similar correction in order to capture flux in the wings of galaxies which is missed by fitting standard light profiles. Using this method, they recover a stellar mass density over thrice that obtained in an earlier estimate with the same sample by \citet{Li09}. However, the \citet{DSou15} value is still only 50\% of that determined by \citet{Bern13}; the characteristic mass is also correspondingly lower, while the slope of the SMF is shallower.

\indent In our fits, we too note that the single Schechter function does not trace the underlying SMF well in these higher stellar mass bins. The difference is particularly marked for the SMF of the disk components where the Schechter fit underestimates the mass density by $\sim$15\% (see Figure \ref{Fig6}, and discussions in \S \ref{SMFbd}). However, we have not adopted the analytic function proposed by \citet{Bern13} for two reasons. In the stellar mass range, 8.9 $\leq$ log$(M/M_{\odot})\leq$11.2, from which we draw our principal inferences, the single Schechter function is an adequate fit to the SMF with the reduced $\chi^2_{\nu}\leq$1.2 in all cases. In addition, the single Schechter function \citep{Sche76} is widely adopted in published literature, therefore comparison of our results with other published results is easier. However, in order to account for any consequent differences in the inferred stellar mass densities between the SMF and the Schechter fit, we provide both values, $\rho_{Sch}$ from the integrated fit, and $\rho_{SMF}$ by numerical integration of the SMF in Table \ref{Tbl2a}. This comparison of our galaxy SMF and stellar mass densities help validate the de Vaucouleurs spheroid and exponential disk light profile fits adopted by $SL11$ as well as their stellar mass estimates given in $MT14$ against corresponding values obtained by \citet{Bern13} with single S\'{e}rsic values. 

\indent The comparison of SMFs and Schechter fits is shown in Figure \ref{Fig7} (\textit{upper panel}), while the differences between the published results and ours are shown in the lower panel. Within the stellar mass range $8.9 \leq$ log($M/M_{\odot}$) $\leq 11.2$, we find that our SMF estimates are largely consistent with the published values within their combined uncertainties ($\Delta$log$(\Phi) \leq 0.1$ dex). This applies to the SMF comparisons with \citet{Bern13} as well as the \citet{Mous13} PRIMUS results, and equally to the double Schechter fit from the GAMA survey \citep{Kelv14a}. Approaching the high mass end, at log($M/M_{\odot}$)$\geq 11.2$ we find that our estimates are higher than those of the PRIMUS and to a greater degree with the GAMA results, which exhibits a sharp downturn. It must be pointed out that the PRIMUS results are higher than both the GAMA estimate as well as our double Schechter fit. As shown in Appendix \ref{IntcompSMFg}, systematic uncertainties begin to dominate our error budget significantly at log($M/M_{\odot}$)$\geq 11.2$, and the increasing differences between published SMF estimates at the high stellar mass end may just be a reflection of this underlying uncertainties in the stellar mass estimates. In this context, it is important to keep in mind the cautionary note raised by \citet{Bern13} on the significant systematic uncertainties in stellar masses arising out of the different fits to the light profiles and the assumptions made in the light-to-mass ratio conversions and SED fitting, especially at the high mass end. The increasing differences between our estimates and these published results highlight the issue and need to be taken into account when using our results in this high stellar mass range.  

\subsection {SMF of the disk and spheroid components of galaxies} \label{SMFbd}

We next obtain the $V_{max}$  weighted SMF of the spheroidal and disk components of this galaxy sample using the same methodology as for the galaxies. The upper right panel of Figure \ref{Fig6} shows the SMF computed using the stellar masses in only the spheroidal components of \textit{all} the galaxies in our catalog. Similarly, the corresponding SMF of the disk components of the complete galaxy sample is shown in the lower left panel. It is important to emphasize that these SMFs labeled \textit{spheroids} and \textit{disks} in Figure \ref{Fig6} reflect the mass functions of the corresponding \textit{galaxy components}, being estimated with the stellar mass in the pressure-supported spheroidal components and that of the rotation-supported disk components of \textit{all} the galaxies in our sample. These are \textit{not} SMFs of disk-dominated and spheroid-dominated sub-populations of galaxies, nor are they those of blue, star forming, and red, passive galaxies, which have been extensively discussed in the literature; we turn our attention to the SMF of spheroid- and disk-\textit{dominated} galaxies, classified using their $(B/T)_*$ values in the next sub-section \ref{SMFbddom}. 

\indent In Figure \ref{Fig6}, the SMFs are plotted as functions of the stellar masses of the corresponding components, and not that of their host galaxies. It is for this reason that the galaxy SMF at any stellar mass is \textit{not} directly equal to the sum of the SMFs of the spheroid and the disk components at that same stellar mass. With the $SL11$ spheroid+disk decomposition, and the $MT14$ catalogs providing the corresponding stellar masses of the spheroidal and the disk components for each galaxy, we are able to independently derive the spheroid and the disk SMFs given here. The uncertainties associated with each SMF due to the quoted uncertainties in the stellar masses in $MT14$ are estimated by applying MC sampling similar to the method used for galaxies. The $16^{th}$ and $84^{th}$ quartile values in each stellar mass bin added in quadrature with the corresponding Poisson uncertainties as well as the systematic uncertainties are used as the total uncertainties in the SMF. These total uncertainties in the SMFs are indicated by the shaded contours in Figure \ref{Fig6}. The Schechter parameters and their associated uncertainties are obtained using methods described in \S \ref{SMFg} for the galaxies, and are listed in Table \ref{Tbl2a}. The fitted Schechter functions are also overplotted on the corresponding SMF for the spheroid and disk components in Figure \ref{Fig6}. The stellar mass densities, $\rho_{SMF}$ and $\rho_{Sch}$ with associated uncertainties for the spheroids and disks are listed in the last two columns of Table \ref{Tbl2a}.

\begin{figure*} 
\includegraphics[scale=0.8]{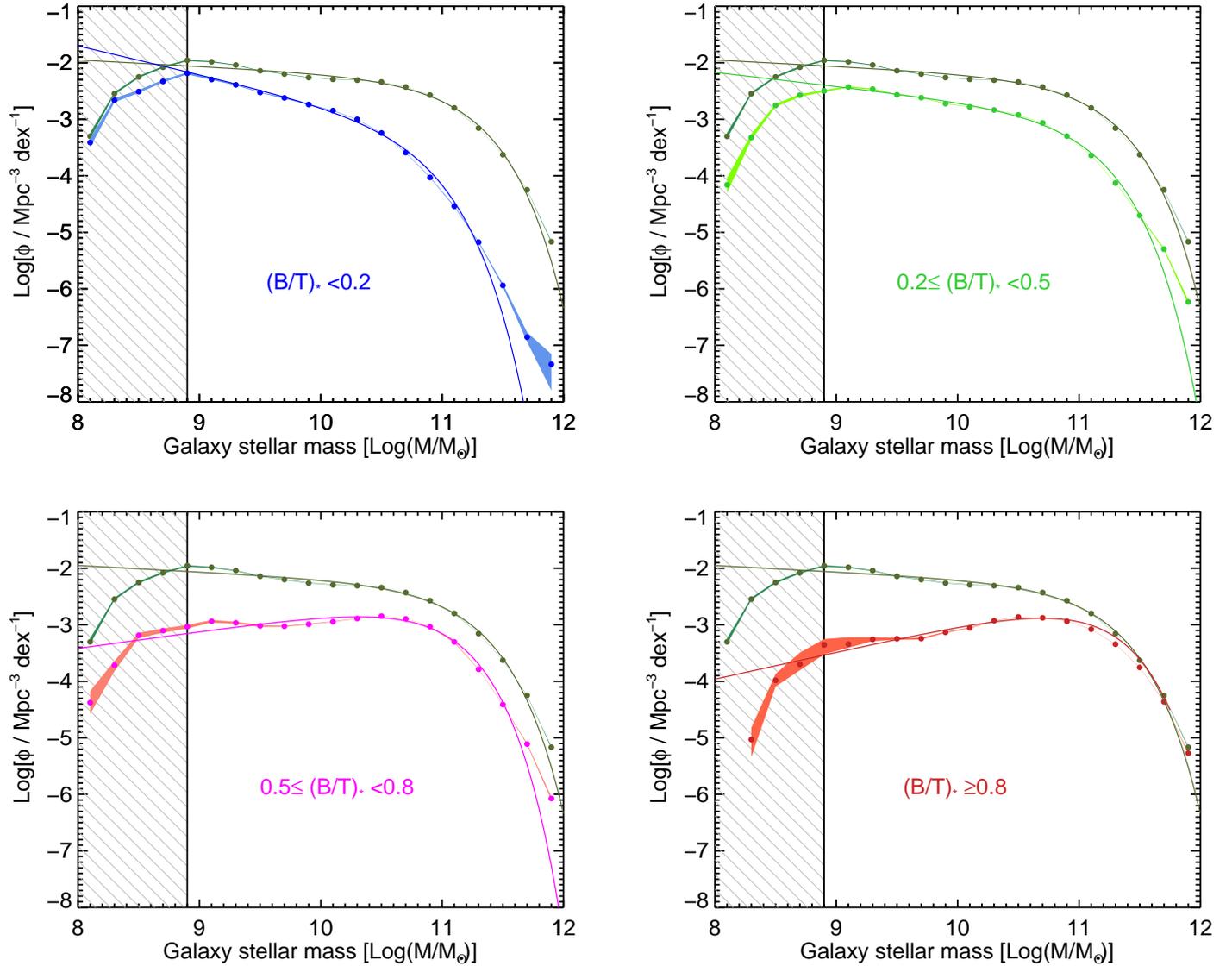}
\caption{Panels showing the SMF of our total galaxy sample compared with that of each of the four sub-groups of galaxies defined in \S \ref{BDclasses} based on their stellar mass B/T values. The single Schechter function fit to the SMF of each sub-popualtion is also overplotted. The corresponding Schechter parameters and the stellar mass densities for the these classes are listed in Table \ref{Tbl3}. (\textit{Color figure available in the online journal})\label{Fig8}}
\end{figure*}

\indent Comparing the SMFs of the spheroidal and disk components of the \textit{complete} galaxy sample, at the high end of the stellar mass range with log$(M/M_{\odot}) \geq$10.2, the stellar mass in spheroids is dominant with their SMF closely tracking that of the galaxies in which they reside; the trend in the disk components on the other hand shows a steep drop off in this higher stellar mass regime (with log($M^*/M_{\odot}$)=11.085$\pm$0.01 for spheroids versus 10.707$\pm$0.007 for disks). However, in the lower stellar mass regions, log(M/$M_{\odot}$) $\leq$10, the stellar mass in the disk components of galaxies are dominant and are more representative of the galaxy SMF while the spheroidal mass function systematically falls $\sim0.5$dex lower than that of the disks and galaxies (with the faint end slope, $\alpha$=-1.277$\pm$0.007 for disks versus -1.073$\pm$0.007 for spheroids). Integrated over the corresponding SMFs, the stellar mass density in the low redshift universe within the stellar mass range, 8.9 $\leq$ log(M/$M_{\odot}$) $\leq$ 12 is seen to be dominated by the stellar mass in spheroids with $\rho_{SMF}\;=1.687^{\textsc{\tiny +0.063}}_{\textsc{\tiny -0.063}}$, while that in the disks is only 54\% of that value at $0.910^{\textsc{\tiny +0.029}}_{\textsc{\tiny -0.029}}$, both in units of $10^8 M_{\odot} Mpc^{-3}$. Extending the mass range to log(M/$M_{\odot}$) $\geq$ 8 to include the more numerous disk components increases their fractional contribution to the mass density only to 55.5\%, indicating that the stellar mass density is dictated primarily by the relative distributions at the high mass end. This further underlines the cautionary note by \citet{Bern13} regarding the importance of properly estimating the stellar mass and the SMF at the high mass end, and fitting an appropriate representative function.    

\indent The two distinct ranges of stellar mass where the SMFs of the spheroidal and disk components dominate are seen in their overplots shown in the lower, right panel of Figure \ref{Fig6}, with the zoomed inset highlighting the stellar mass where the SMFs intersect. We refer to this as the \textit{crossover stellar mass}. Since the SMF of the spheroids and disks are generated based on the stellar mass in each of these two components, this crossover stellar mass, which lies at their intersection, represents the stellar mass at which the volume weighted space density of the spheroidal component of galaxies equals that of disk components. The possible implications of the crossover stellar mass and its possible utility for comparison with hydrodynamical models are discussed in \S \ref{Disc}. 

\indent To locate the point of intersection, we set the two Schechter functions equal and use the Broyden method \citep{Pres92} to solve the resulting nonlinear equation. The crossover mass is found to be log(M/$M_{\odot}$) = 10.3$\pm$0.030. The quoted mean crossover mass value and associated uncertainty have been derived using MC sampling in which we let the fit uncertainties in the Schechter parameters, $\rho$, $M^*$ and $\alpha$ for the spheroids and disks to range fully over the corresponding uncertainty values listed in Table \ref{Tbl2a}. Because the crossover mass occurs in the middle of our stellar mass range, it is reasonably unaffected by the incompleteness of our sample in the lower stellar mass regions, and by the systematic uncertainties which affect the higher mass bins.  


\subsection {SMF of spheroid and disk dominated systems} \label{SMFbddom}

With this access to a wealth of structural and stellar mass properties, next we estimate the SMF of the spheroid- and the disk-dominated sub-populations, and thus evaluate their individual contributions to the overall galaxy SMF. For this, we bin the galaxies into four sub-groups, listed in Table \ref{Tbl1}, based on their stellar mass $(B/T)_*$ values. 

 \begin{table*}
 \caption{Schechter parameters for spheroid- and disk-dominated galaxies}
 \label{Tbl3}
 \begin{center}
 \scriptsize
 {\renewcommand{\arraystretch}{2.0}
  \begin{tabular}{|L{2cm}|C{3cm}|C{2cm}|C{2cm}|C{2cm}|C{2cm}|}
 \hline
 \hline
 Population & $\phi^*$ & $log_{10}\;(M^*/M_{\odot})$ & $\alpha$ & $\rho_{Sch}$ & $\mathbf{\rho_{SMF}}$ \\
  \hline
  & [$10^{-3}\; Mpc^{-3}\; dex^{-1}$] & [ - ] & [ - ] & [$10^8\;M_{\odot}\;Mpc^{-3}$] & [$10^8\;M_{\odot}\;Mpc^{-3}$] \\
\hline
B/T $\leq$ 0.2 & $0.332^{+0.013}_{-0.013}$ & $10.740^{+0.010}_{-0.010}$ & $-1.524^{+0.013}_{-0.013}$ & $0.320^{+0.020}_{-0.020}$ & $0.287^{+0.020}_{-0.020}$ \\ 
 \hline
0.2 $\leq$ B/T $<$ 0.5 & $0.489^{+0.015}_{-0.014}$ & $10.988^{+0.009}_{-0.010}$ & $-1.270^{+0.010}_{-0.010}$ & $0.592^{+0.027}_{-0.027}$ & $0.560^{+0.027}_{-0.027}$ \\ 
 \hline
0.5 $\leq$ B/T $<$ 0.8 & $1.016^{+0.023}_{-0.022}$ & $10.926^{+0.009}_{-0.009}$ & $-0.764^{+0.015}_{-0.014}$ & $0.779^{+0.026}_{-0.026}$ & $0.717^{+0.026}_{-0.027}$ \\ 
 \hline
B/T $\geq$ 0.8 & $1.203^{+0.028}_{-0.026}$ & $11.023^{+0.011}_{-0.011}$ & $-0.580^{+0.018}_{-0.017}$ & $1.124^{+0.041}_{-0.041}$ & $1.055^{+0.041}_{-0.042}$ \\ 
 \hline

\end{tabular}
}
 \end{center}
 \begin{flushleft}
 {\footnotesize Fitted Schechter parameters describing the SMF of the SDSS-DR7 population of galaxies split by their stellar mass B/T into four sub-groups, as described in \S \ref{BDclasses}.}
 \end{flushleft}
 
 \end{table*}
 \normalsize 

\indent The SMFs of these four sub-groups of galaxies are shown in the four panels in Figure \ref{Fig8} with each compared with the overall galaxy SMF to visually highlight the relative contribution of each sub-group to the total; as expected, the SMFs of these four galaxy subpopulations add up to the overall galaxy SMF within their combined uncertainties. The associated uncertainties for each SMF have been estimated using MC sampling as described in \S \ref{SMFbd}, and the width of the filled regions in Figure \ref{Fig8} represent these uncertainties as a function of stellar mass. For estimating the corresponding Schechter parameters for each class, we fit the single Schechter function to the 100 MC realizations of the SMF and the corresponding uncertainties of that class. The Schechter parameters and their uncertainties listed in Table \ref{Tbl3} represent respectively the median and the $16^{th}$ and $84^{th}$ quartiles obtained from this MC sampling. The last two columns in Table \ref{Tbl3} list the stellar mass densities, $\rho_{Sch}$ and $\rho_{SMF}$ for each class, and are obtained as described in \S \ref{SMFbd}.

\indent The single Schechter function is a reasonable representation to the SMF of these sub-populations of galaxies in the stellar mass range, $8.9 \leq log(M/M_{\odot}) \leq 11$. However, the fit becomes progressively poor outside this stellar mass range, especially at the high mass end for both the spheroid-dominated $BT100$, and the disk-dominated $BT20$ classes - the Schechter fit overpredicts the contribution of the $BT100$ galaxies to the mass function, while the steep drop off in this mass range for the $BT20$ class results in an underestimate. However, for consistency we fit the Schechter functions for these sub-populations over the same range used for the overall galaxy sample, $8.9 \leq log(M/M_{\odot}) \leq 12$. In light of the cautionary remarks of \citet{Bern13} regarding this, in \S \ref{Disc} we return to a discussion of possible consequences, and our plans to address them in a forthcoming publication.

\indent The shift seen in the trends in the SMF as we progress from the low to high $(B/T)_*$ classes is evident in the four panels in Figure \ref{Fig8}. The disk-dominated $BT20$ population has a steep faint end slope ($\alpha = -1.524\pm0.013$), with a sharp drop off at the high mass end (log($M^*/M_{\odot}$)=10.740$\pm$0.01). On the other hand, the spheroid-dominated $BT100$ population shows the inverse trend, contributing almost the entirety of the overall galaxy SMF at the high mass end, with $log(M^*/M_{\odot})=11.023\pm0.011$; there is however a sharp downturn at the lower stellar mass bins with $\alpha = -0.58\pm0.018$. These classes perhaps represent the two distinct classes that populate the \textit{blue cloud} and the \textit{red sequence} in the stellar mass versus ($g-r$) color plane, as seen in Figure \ref{Fig4} and more so in Figure \ref{Fig5}. The dominance of the high $(B/T)_*$ population at the high stellar mass end of the color distribution seen in those plots is reflected directly in the SMF comparison as well. The SMF of the disk- and spheroid-dominated composite galaxies fall in between these two extremes, with the spheroid-dominated $BT80$ galaxies contributing $\sim$0.6 dex more than their disk-dominated $BT50$ counterparts in the higher stellar mass bins, with log$(M^*/M_{\odot})=10.988\pm0.01$ versus $10.926\pm0.009$ respectively. At the low mass end however, the trends reverse as shown by faint end slope, with $\alpha=-0.764\pm0.015$ and $-1.27\pm0.01$ for the two classes respectively, and the disk-dominated galaxies become the principal contributors to the overall galaxy SMF.     

\begin{figure*} 
\includegraphics[scale=0.65]{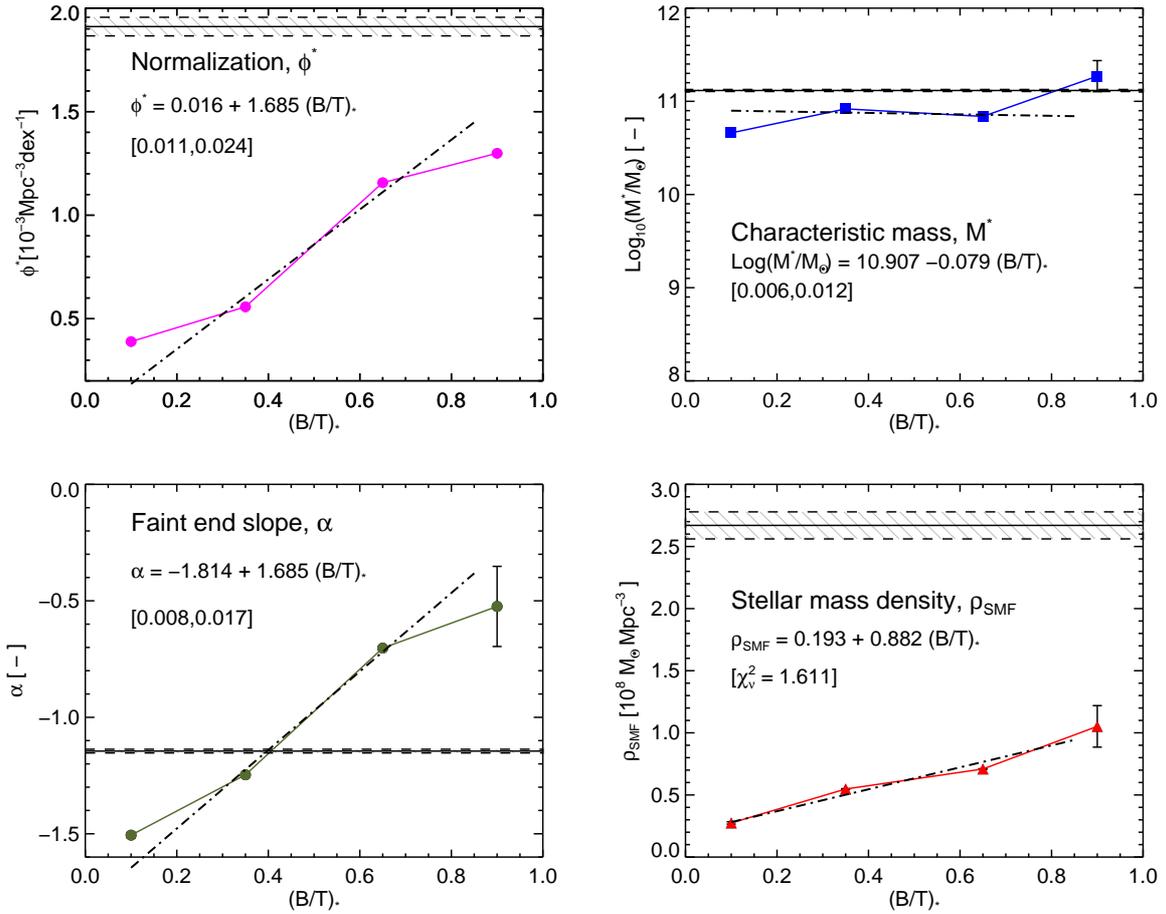}
\caption{Trends in the Schechter parameters fitted to the SMFs of the four galaxy sub-populations classified by their stellar mass, $(B/T)_*$ ratio. Each parameter is compared to its corresponding value for the total galaxy population, which is shown by the solid line with the hatched region indicating the associated uncertainty. Also overplotted is a linear fit representing the trend in each parameter as  a function of the $(B/T)_*$ ratio; the corresponding linear relation and the associated formal uncertainties in the estimated parameters are also annotated on each panel. (\textit{Color figure available in the online journal}) \label{Fig9}}
\end{figure*}

\indent The stellar mass density, $\rho_{SMF}$ given in the last column in Table \ref{Tbl3} for these classes running from the highest to lowest $(B/T)_*$ values are ($1.055\pm0.041$), ($0.717\pm0.027$), ($0.560\pm0.027$) and ($0.287\pm0.02$) respectively, in units of $10^8 M_{\odot} Mpc^{-3}$. The sum of the stellar mass densities of the four classes again agrees with the total galaxy stellar mass density well within a combined $1\sigma$ uncertainty in their estimates. Even though the disk-dominated galaxies contribute significantly more than their spheroid-dominated counterparts to the volume weighted galaxy \textit{number densities} at lower stellar masses as indicated by their steeply rising faint end slopes, the spheroid-dominated galaxies contribute over two-thirds of the overall galaxy stellar mass density due to their dominance at the high stellar mass end. The increasing numbers of spheroid-dominated galaxies in the higher mass bins is amplified by the over two orders of magnitude increase in the mean stellar mass per galaxy, which therefore results in this significant difference in their contributions to the mass density.

\indent In order to highlight these trends in the SMFs of the $(B/T)_*$ based galaxy sub-populations, in the four panels in Figure \ref{Fig9} we plot the values of each Schechter parameter and the stellar mass density against the $(B/T)_*$ values of the four classes. For comparison, the value of the corresponding Schechter parameter for the overall galaxy population as listed in Table \ref{Tbl2a} is also shown on each plot as a line with the hatched region representing the associated uncertainty. In order to quantify the trend, we have used a simple linear fit to the parameter values and associated uncertainties as a function of $(B/T)_*$ ratio. Annotated on each panel is the functional form of the linear fit. The formal 1$\sigma$ uncertainties in the fit values are given in square brackets for the three Schechter parameters, while the reduced $\chi^2_\nu$ value is given for the stellar mass density. 

\indent The plotted trends indicate a monotonic increase in the characteristic mass with $(B/T)_*$, while correspondingly the faint end slope becomes progressively less steep. The increase in the normalization constant with $(B/T)_*$ reflects the increasing contributions of the spheroid-dominated galaxies to the overall stellar mass budget as shown by the corresponding rise in the $\rho_{SMF}$ values. In \S \ref{Disc}, we discuss possible implications of these trends and their effects on the underlying SMFs. In a follow-up publication, we aim to explore these correlations more systematically with a more representative fitting function. Here, we next compare our SMF estimates for these four classes against two recent published results, those of the $E/S0$ and \textit{Spiral} classes provided by the GAMA survey \citep{Kelv14a}, and by \citet{Mous13} for PRIMUS for the corresponding \textit{Quiescent} and \textit{Star Forming} galaxy classes defined by them. 

\subsection{External consistency checks: Disk- and Spheroid-dominated SMF} \label{ExtcompSMFds}

\begin{figure*} 
\includegraphics[width=0.45\textwidth]{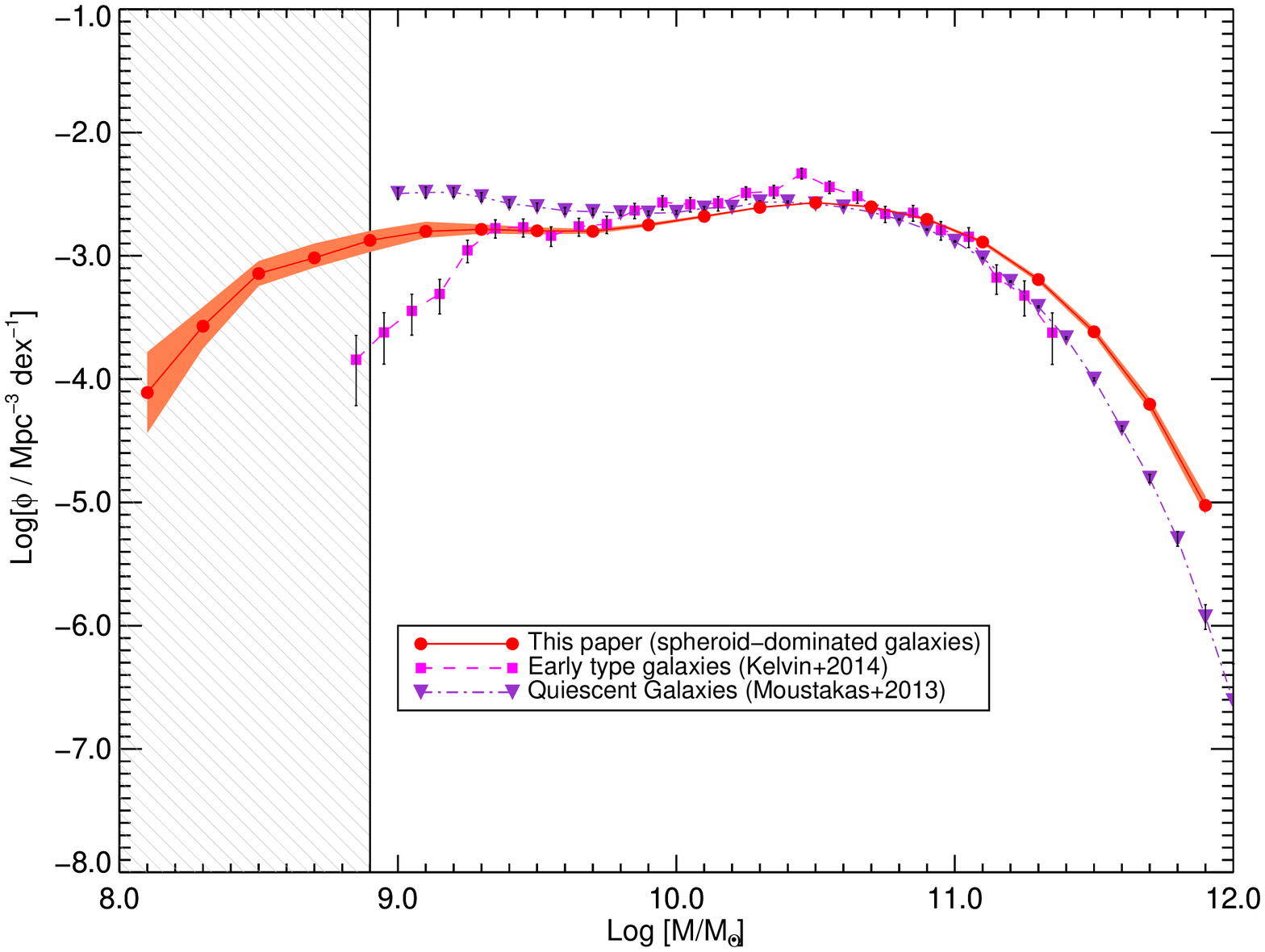}
\includegraphics[width=0.45\textwidth]{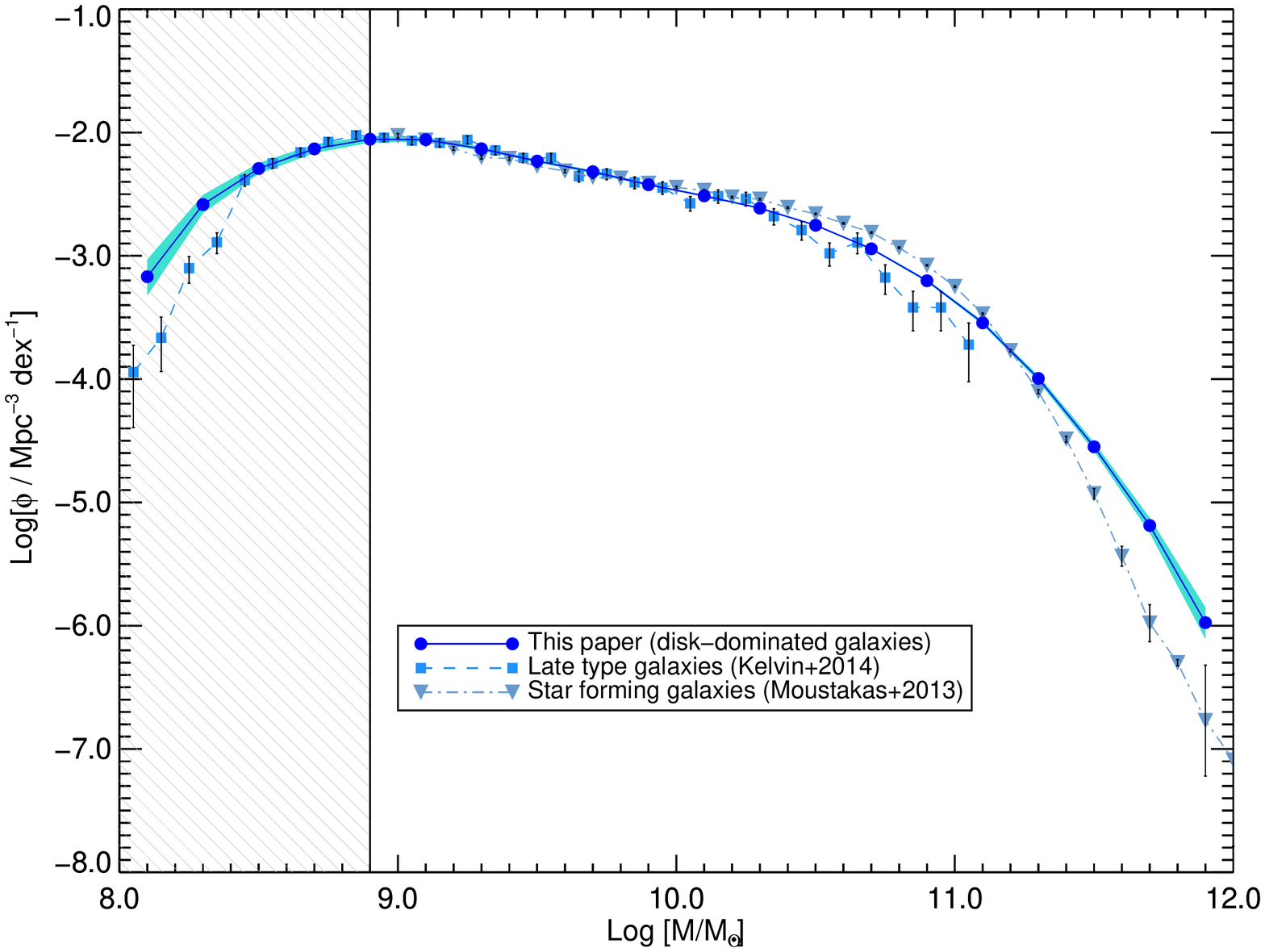}
\caption{Comparisons of our SMF for the spheroid-dominated ($B/T \geq 0.5$) galaxies (\textit{left panel}), and the disk-dominated ($B/T < 0.5$) galaxy population with corresponding SMFs for the quiescent, early-type galaxies and the late-type, \textit{star forming} galaxies in the GAMA \citep{Kelv14a}  and the PRIMUS \citep{Mous13} surveys. The width of the shaded region and the error bars represent 1$\sigma$ uncertainties in the corresponding values. (\textit{Color figure available in the online journal}) \label{Fig10}}
\end{figure*}

\citet{Mous13} and \citet{Kelv14a} adopt different classification methods for the PRIMUS and GAMA galaxy samples from which they estimate the SMF of the disk- and spheroid-dominated galaxies for each survey. The GAMA galaxies are classified by visual inspection by three expert classifiers into the principal Hubble types, ellipticals, lenticulars, spirals and irregulars, with the late type galaxies being further subdivided into barred and unbarred versions; \citet{Kelv14b} provide full details of the morphological types used and the rules adopted for their classification scheme. On the other hand, \citet{Mous13} use the multi-wavelength photometry as well as spectroscopic data available for the PRIMUS survey to split their sample into quiescent and star forming galaxies based on whether they lie above or below the \textit{star forming sequence} \citep{Noes07}. For our SDSS-DR7 sample, we use the structural parameters from GIM2D as given in the $SL11$ catalogs, and the stellar masses and corresponding $(B/T)_*$ values from $MT14$ to categorize the galaxies into four classes based on their disk and spheroid dominance, as explained in \S \ref{SMFbddom}. However, given these differing classification schemes, in order to keep this comparison simple and fair, we split our sample into just two classes, spheroid- and disk-dominated galaxies with $(B/T)_* \geq 0.5$ being \textit{spheroid-dominated}, and $< 0.5$ being \textit{disk-dominated} in the discussions below.

\indent The panels in Figure \ref{Fig10} show the comparisons of the SMFs of the spheroid- and disk-dominated galaxies against those for the PRIMUS and GAMA samples. The SMF values for the early- and late-type galaxies in the GAMA survey were provided by L.~S.~Kelvin (\textit{priv. comm.}) For the PRIMUS galaxies, the SMFs are provided in tabulated form in their Table 3, \citet{Mous13}. Given the poor fit of the Schechter function to the SMFs of these galaxy sub-populations at the high and low stellar mass ends of our range, we do not include the fitted functions in this comparison.

\indent In the case of the disk-dominated galaxies, our estimates track those of the PRIMUS results well within the combined uncertainties throughout the stellar mass range of interest despite the different classification schemes used. However, for this galaxy class, our results fall below those of \citet{Kelv14a} for the GAMA sample throughout the same stellar mass range. In the case of the spheroid-dominated class, the match with the \citet{Mous13} results for the PRIMUS survey are poorer, with our estimate being greater at the high mass end, and smaller at the low mass end. In the case of the GAMA results for the spheroid-dominated class, the match is within the combined uncertainties only in the region around the characteristic stellar mass; however, our SMF values are higher than the \citet{Kelv14a} values at both the high and low mass ends. Given this match with the PRIMUS results for the disk and to a lesser extend to the spheroid dominated sub-groups, our stellar mass $(B/T)_*$ based classification appears to match their star forming sequence (color-magnitude) based bimodal segregation of the galaxy population and their characteristics. On the other hand, the difference with the GAMA results may be indicative of uncertainties in the GIM2D disk+spheroid decompositions, as well as the challenges of visual classification of spheroidal and lenticular systems, and the issues of reddening and consequent contamination, which are discussed further in \S \ref{Disc}.        

\subsection {Relative contributions of the disk and spheroid to galaxy stellar mass} \label{bd2gSM}

\begin{figure*} 
\includegraphics[scale=0.7]{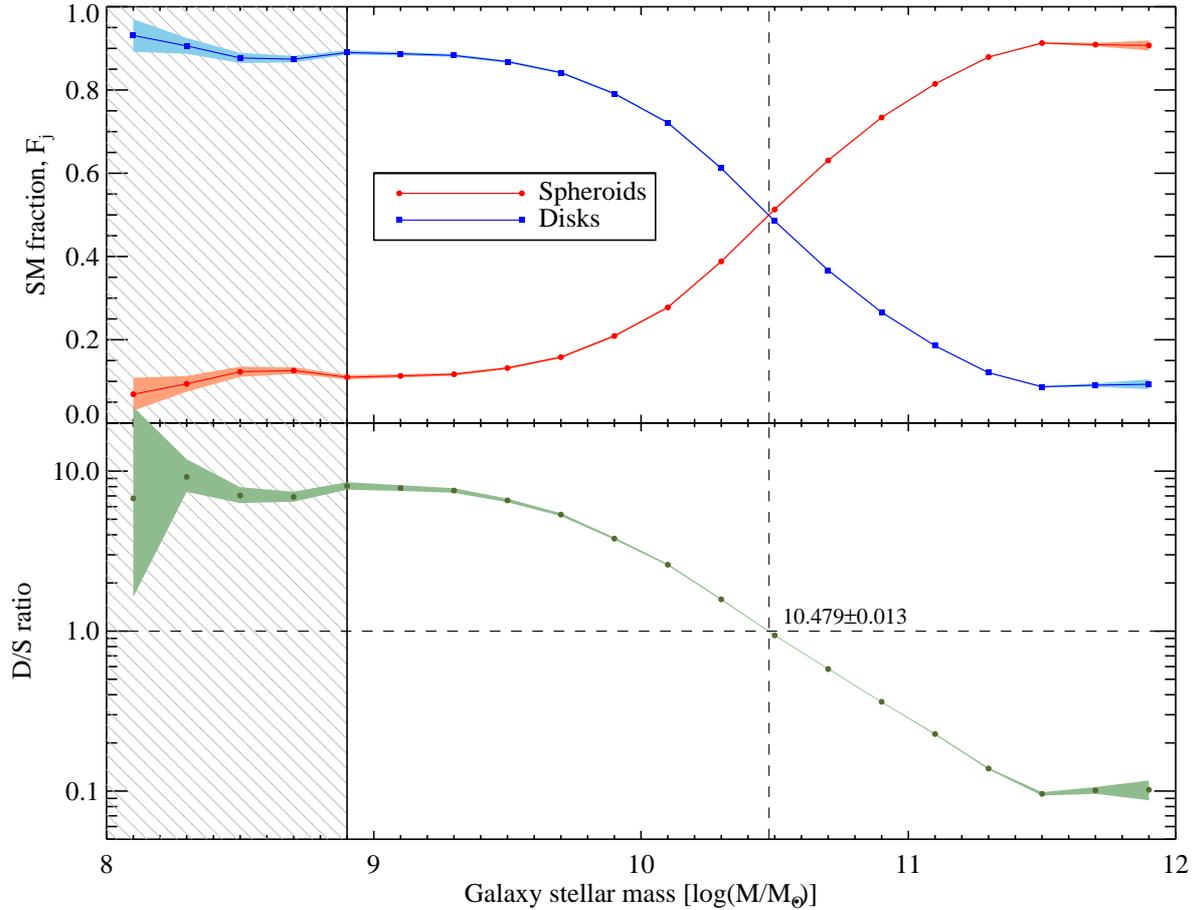}
\caption{Relative contributions of the disk and spheroidal components to the stellar mass budget in galaxies binned by stellar mass. (\emph{Top panel}) Relative contribution of each component shown individually; (\emph{bottom panel}) relative fraction of the stellar mass in the disk over that of the spheroid as a function of galaxy stellar mass. The contributions are equal at a galaxy stellar mass of 10.479$\pm$0.013 log$(M/M_{\odot})$, shown annotated. (\textit{Color figure available in the online journal}) \label{Fig11}}
\end{figure*}

Along with the SMFs of the low redshift disk and spheroidal populations, a closely allied question is the relative fractional contributions of these two components to the stellar mass budget of their host galaxies, and how this fraction varies with stellar mass. Using a sample of 600 galaxies, \citet{Bens02b} estimated that the disk-to-spheroid stellar mass ratio in the local universe is 1.3 for galaxies with stellar mass, $M \geq 10^9M_{\odot}$, indicating that disks contribute 30\% more to the stellar mass budget than spheroids in galaxies in the low redshift Universe. With a bigger sample of 10095 galaxies from the Millennium Galaxy catalog, \citet{Driv07} obtain a mass density breakdown of 10\% in elliptical galaxies, 29\% in galaxy bulges, 58\% in disks and 3\% in other classes (for ease of comparison, we combine all the non-disk contributions to yield 42\% in spheroids). It is important to point out that even though \citet{Driv07} too use GIM2D for the bugle+disk decomposition of their galaxy sample, their $(B/T)$-based morphology classification is based on the measured $r$-band \textit{flux} and not on the stellar mass in the galaxy components, which we use. Our choice is based on the results presented by \citet{Bluc14} who discuss the appreciable differences in the properties of the resulting sub-populations from using a single-band flux versus the stellar mass from SED fitting as the classification parameter.

\indent The significant size of our SDSS DR7 + GIM2D galaxy catalog provides several thousands to tens of thousands of galaxies in each unit log stellar mass bin in the range $8.9 \leq \mathrm{log}\;M \leq 12$, and is thus well suited to not only verify these earlier observational results, but also permits us to assess the variation of the spheroid-to-disk stellar mass ratio as a function of the galaxy stellar mass. To obtain the relative contributions of the spheroid and disk components to the stellar mass of the host galaxy, we split the galaxy sample into 0.2 log stellar mass bins, and obtain the fractions of the stellar masses residing in disks and spheroids relative to the galaxy stellar mass (summed over all the galaxies in each stellar mass bin). In particular, the stellar mass fraction in spheroids, $F_j$, in each galaxy stellar mass bin is expressed as,
\begin{equation}
\label{SMfrac}
F_j = \frac{\sum_{i=1}^{n_j}\,(w_i/V_{max,i})\mathrm{M}_{(\mathrm{spheroid}), i}}{\sum_{i=1}^{n_j}\,(w_i/V_{max,i})\mathrm{M}_{(\mathrm{galaxy}), i}}
\end{equation}
 where, $n_j$ is the number of galaxies in that stellar mass bin, with $j$ being the bin index. Since we do not include any correction for $TSR$, as explained in \S \ref{SMFg}, the weighting function, $w_i = 1$ in the above equation. The corresponding disk stellar mass fraction is obtained using a similar expression. The trends in these relative fractions as a function of galaxy stellar mass are shown for the disk and spheroid in the upper panel of Figure \ref{Fig11}; the width of each trace is representative of the 1$\sigma$ uncertainty in the relative fraction determined using the jackknife technique. The fraction of the disk stellar mass relative to that in spheroids in each stellar mass bin is shown in the bottom panel. 
 
 \indent Figure \ref{Fig11} shows that in the sub-$M^*$ galaxies, the bulk of stellar mass is in disks, with the summed disk mass being $~5\times$ that of spheroids in galaxies of stellar mass $\sim 10^9 M_{\odot}$. On average, these disk-dominated galaxies have $(B/T)_*$ values $\leq0.25$. However, with increasing galaxy stellar mass, this disk dominance gradually decreases with a corresponding increase in the spheroidal contribution. This trend leads to each component contributing equally in galaxies with stellar mass, log$(M/M_{\odot})= 10.479\pm0.013$, determined by interpolation in Figure \ref{Fig11}; the associated uncertainty is estimated by letting the fractional contributions range fully over their associated uncertainties in a Monte Carlo sense using $10^4$ trials, then determining the median and the interquartile distance of the resulting set of intersection points. Our estimate of this ratio indicates that the equal contribution of the spheroid and disk to the galaxy stellar mass estimated by \citet{Bens02b} applies more to Milky-way type galaxies with stellar mass, $M \sim 10^{10.3}M_{\odot}$. In more massive galaxies, spheroids contribute an increasing fraction of the host galaxy's stellar mass, with the stellar mass in spheroids increasing to $\sim9\times$ the mass in disks in galaxies of stellar mass, $M \geq 10^{11.5}\,M_{\odot}$. In comparison with the \citet{Driv07} results, we estimate the overall contribution from the spheroidal components to the galaxy stellar mass to be much higher, $\sim65$\%. In \S \ref{Disc}, we discuss the implications of these results in the light of theoretical predictions and hydrodynamical simulations.

\subsection {Census of disk and spheroid dominant galaxies} \label{Nbd2gSM}

\begin{figure*} 
\includegraphics[scale=0.8]{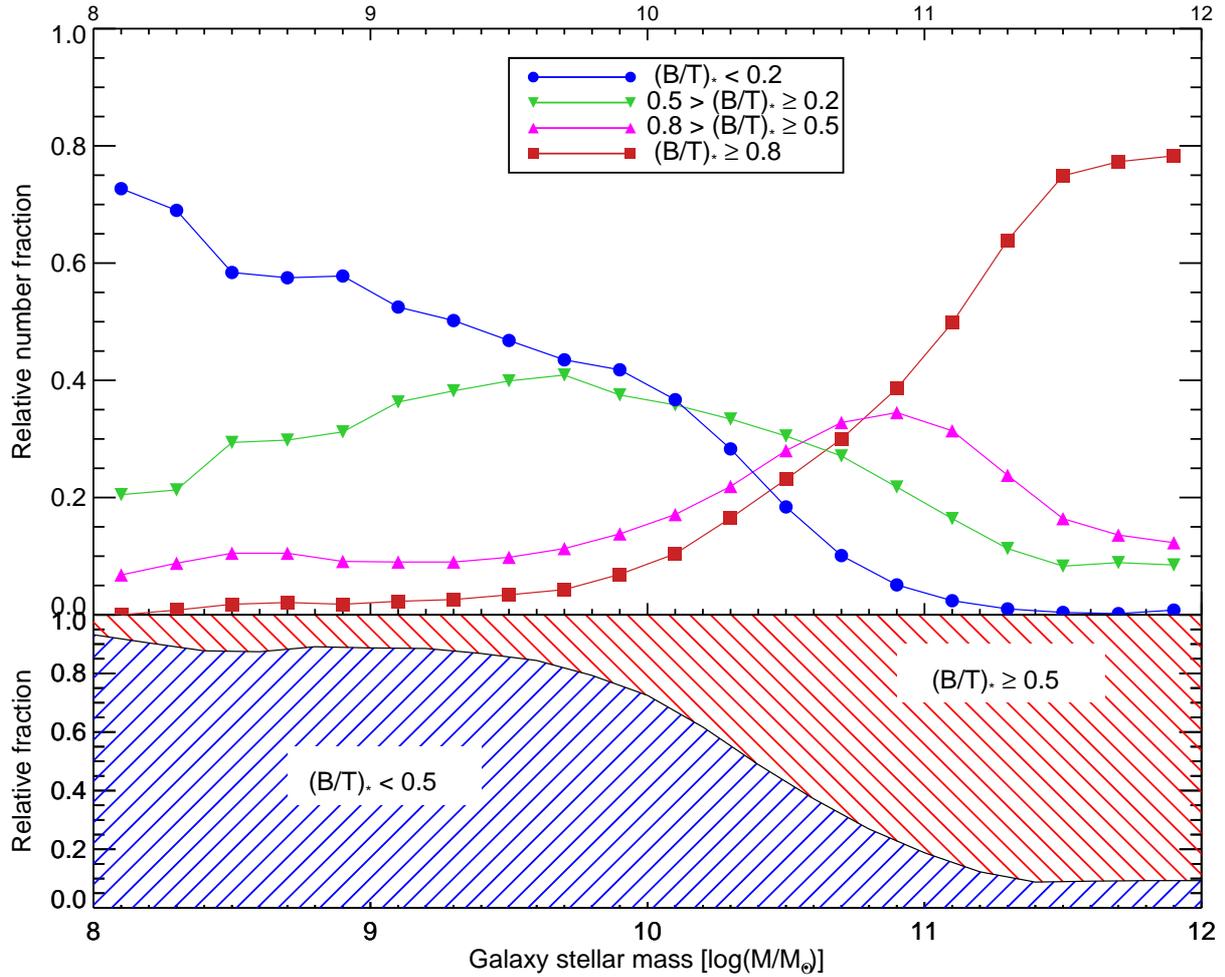}
\caption{(\textit{Upper}) The distribution of the relative numbers of single component and composite disk dominated galaxies, and their corresponding spheroid dominant counterparts computed as a fraction of the number of galaxies in each $0.2\, \mathrm{log}\,M_{\odot}$ galaxy stellar mass bin; the galaxy structural classification is defined in \S \ref{BDclasses}. (\textit{Lower}) A \textit{ying-yang} plot showing the relative number of disk versus spheroid dominated galaxies in each 0.2 log$M_{\odot}$ bin. (\textit{Color figure available in the online journal}) \label{Fig12}}
\end{figure*}

Along with the relative contributions of the disk and spheroid components to the stellar mass of galaxies in the local universe, the SDSS-DR7 sample permits us to estimate the relative \textit{numbers} of disk- and spheroid-dominated galaxies in the galaxy population binned by stellar mass in the stellar mass range 8.9$\leq$log($M/M_{\odot}$)$\leq$12. In order to distinguish between the dominance of these structural components, we retain the classification based on stellar mass $(B/T)_*$ values defined in \S \ref{BDclasses}. 
 
\indent After binning the overall galaxy population in 0.2 dex bins, in Figure \ref{Fig12} (\textit{upper}) we trace the relative numbers of galaxies in each of the four classes in each stellar mass bin, taken as a fraction of the total number of galaxies in that bin. The fractional contribution of each class to the number of galaxies in each stellar mass bin may be directly read off the y-scale, with their sum being equal to one. In the \textit{ying-yang} plot in the lower panel, we compare the relative numbers of the disk-dominated (with $B/T < 0.5$) versus spheroid-dominated galaxies as a function of the galaxy stellar mass; the color coded hatched areas highlight the contribution of each class. 

\indent From this galaxy census, it is seen that over $85\%$ of the population below log($M/M_{\odot}$)$ \leq$ 9.5 is made up of low $(B/T)_*$, disk-dominated galaxies, with a sizeable fraction of the rest provided by their composite counterparts with mean $(B/T)_* \sim0.25$. Taken together they account for over $95\%$ of the number of galaxies in this stellar mass range. With increasing stellar mass, this disk dominance in galaxy number counts diminishes rapidly as we approach the crossover stellar mass. Correspondingly, the number of galaxies with significant spheroids increases so that at the crossover galaxy stellar mass of $M \sim 10^{10.5}\;M_{\odot}$ determined in \S \ref{bd2gSM}, the population is made up of equal numbers of spheroid- and disk-dominated galaxies. Further, the number of galaxies in this bin is split almost equally amongst the four classes. Approaching the high mass end with log($M/M_{\odot}$)$\geq$11.5, the galaxy population is dominated ($\geq 80\%$) by the spheroidal galaxies with $B/T \geq 0.8$. The composite spheroids and disks make up the remaining numbers in equal proportions, while the pure disks are virtually absent. We discuss the implications of our findings in the light of theoretical predictions in the following section. 


\section {Discussion} \label{Disc}

\citet{Kelv14a} have argued that the stellar masses of a galaxy sample combined with the morphology or parametrization of their light profiles may define all their other observed and estimated characteristics. As an illustration, they present the bivariate distribution of stellar mass and ($g-i$) rest frame color classified by the various morphological types used for their GAMA sample (see Figure 5 in \citet{Kelv14a}). In this distribution, the \textit{red sequence} defined mainly by the spheroidal $E$-type galaxies shows an increasing scatter at the low mass end of the sequence (where log($M/M_{\odot}$)$\leq$10.2). Two reasons set forth to explain this scatter are (i) face-on, lenticular $S0$ galaxies being misclassified as spheroidal galaxies, and (ii) contamination from \textit{edge-on}, disk-dominated $(Sab-Scd)$ and $(Sd-Irr)$ classes reddened due to dust in their disks. Regarding the misclassification, it must be mentioned that the morphological typing of the GAMA galaxies is done by visual inspection by three experts; the challenges of visually classifying galaxies and especially that of quantifying any selection bias have been discussed in literature, for example, by \citet{Bamf09} and \citet{Lint08} in the context of the \textit{Galaxy Zoo} project. Similar challenges and uncertainties also apply to automated profile fitters such as GIM2D, even though the selection bias may be better quantified. 

\indent Given these challenges of morphological typing of galaxies, we posit that the stellar mass $(B/T)_*$ ratio when combined with the galaxy stellar mass may provide an alternate, quantifiable discriminant of the characteristics of the galaxies. Unlike the discrete galaxy types, the bulge-to-total stellar mass ratio, $(B/T)_*$ is a continuously varying function, which makes tracing trends in various characteristics, and interpolation for intermediate values feasible. At the same time however, we admit that this ratio is based on stellar mass estimates of galaxies and of their spheroid and disk components, and hence is only as good as the underlying spheroid+disk decomposition and the stellar mass estimates. Simplifying assumptions made in these crucial steps will be directly reflected in the uncertainties associated with any inferences drawn from these characterizations.

 \indent The availability of the $SL11$ public catalogs with spheroid+disk decomposition of a large sample of galaxies from the SDSS DR7 spectroscopic catalogs, plus the stellar masses for the galaxies and of their spheroid and disk components in $MT14$ catalogs make possible our adopted classification strategy of galaxies by stellar mass and $(B/T)_*$, and thus study their correlations with various characteristics. As an illustration, in \S \ref{ClrMig} (Figure \ref{Fig4}) we present a comparative bivariate stellar mass and ($g-r$) rest frame color distribution, classified by $(B/T)_*$ instead of morphology. In the distribution of the $BT100$ class (lower, right panel), which closely corresponds to the $E$-type galaxies in \citet{Kelv14a}, over 95\% of this sub-population comprised of $\sim$200,000 galaxies lies along a well defined \textit{red sequence} as shown by the overlaid contours. Similarly, an equal fraction of the disk-dominated $BT20$ class (akin to the $(Sd-Irr)$ class of \citet{Kelv14a}) is restricted to the \textit{blue cloud}. Even this coarse $(B/T)_*$ binning permits us to trace the observed migration of galaxies through the \textit{green valley} in the intermediate $(B/T)_*$ bins. This is highlighted further in Figure \ref{Fig5} where the stellar mass at which this redward trend sets in for each $(B/T)_*$ bin is shown by the location of the knees in four plots. For galaxies in higher $(B/T)_*$ classes, the trend toward the \textit{red sequence} sets in at lower stellar masses.           

\indent Regarding the galaxy stellar mass density in the low redshift universe, \citet{Mous13} find $\rho_{SMF} = 2.36\times10^8\;M_{\odot} \mathrm{Mpc}^{-3}$ for the stellar mass range, $9 \leq log(M/M_{\odot}) \leq 12$ based on the PRIMUS survey. Of this, they estimate $\sim$60\% resides in quiescent galaxies, with the remaining in the active, star-forming, blue population. By comparison, for the GAMA sample, the \citet{Kelv14a} estimate of stellar mass density by morphological type in the low redhift universe puts $71\pm4$\% in spheroid-dominated $(E-S0)$ galaxies, with the remaining $29\pm4$\% in late type $(Sab-Scd)$ and $(Sd-Irr)$ classes. Using $B/T$ values from \citet{Grah08} for various morphological classes, they convert the stellar mass densities of these galaxy sub-populations to those of their spheroid and disk components, and find that their contributions are nearly equal in the low redshift universe. 

\indent Based on our results presented in \S \ref{SMFbd}, we find a higher fractional contribution from the spheroidal components of galaxies to the galaxy stellar mass density. Of the total galaxy stellar mass density of $2.67\pm0.11 \times 10^8 M_{\odot} \mathrm{Mpc}^{-3}$, the stellar mass in their spheroidal components is $\sim$65\%, putting our estimate 15\% higher than the value determined by \citet{Kelv14a}. The cause for this difference is seen in Figure \ref{Fig10} where our estimate of the SMF for the spheroid-dominated galaxies is $\sim1$ dex higher than that of \citet{Kelv14a} especially at the higher stellar mass bins from which the bulk of the contribution to the stellar mass density comes. Classifying the population based on their $(B/T)_*$ ratio, we find that the spheroid-dominated classes (with $(B/T)_* \geq 50$\%) contribute $1.772\pm0.035$ ($\sim$68\%), closely matching the results from the recent PRIMUS \citep{Mous13} and GAMA results \citep{Kelv14a}, but over 20\% higher than the value determined by \citet{Driv07}; properly constraining the SMF at the high mass end appears to be key, in keeping with the cautionary note by \citet{Bern13}.        

\indent Even though these comparisons with recent observational results are useful for determining the strengths and associated uncertainties of our results, our intended audiences are the ongoing large numerical simulations such as the \textit{Illustris} \citep{Voge14a}, \textit{Eagles} \citep{Scha15} and the \textit{FIRE} \citep{Hopk14} projects. These projects generate extensive synthetic galaxy catalogs from which one may directly estimate the individual contributions of the stellar masses in the spheroid and disk components, without introducing uncertainties from the decomposition of the observed light profiles. By making available the SMFs and the stellar mass densities for the spheroidal and the disk components of a large observed galaxy sample, our goal is to provide direct observational constraints for these suites of numerical simulations. By combining the SMFs of the spheroid and disk components, the \textit{crossover stellar mass} in Figure \ref{Fig6} is intended to act as an even more stringent observational constraint. Similar comments apply to the SMF and stellar mass densities of the galaxy sub-populations classified by their $(B/T)_*$ values provided in \S \ref{SMFbddom}. This quantitative classification metric may be directly applied to the synthetic galaxy catalogs, and their stellar mass characteristics compared with our observational results.
    
\indent In the build up of stellar mass in the spheroidal and disk components as a function of their host galaxy stellar mass, theoretical predictions assign disks as the principal contributors in sub-$M^*$ galaxies while galaxies with stellar mass $\geq 10^{11}\,M_{\odot}$ are spheroid dominated \citep{Dutt09, Stew08, Stew09a}. Given the predicted major merger histories of these massive galaxies, \citet{Dutt09} draws the stronger conclusion that the bulge fraction of galaxies is a strong function of the stellar mass - in sub-$M^*$ galaxies ($M\sim 10^{10} M_{\odot}$), over 45\% do not have any significant contribution from the spheroidal component. However the fractional contribution from spheroids increases steeply with stellar mass so that in galaxies more massive than the Milky Way ($M > 10^{11}M_{\odot}$), only a very small fraction ($\ll 1$)\% of the population are bulgeless. 

\indent In \S \ref{bd2gSM}, our estimate of the crossover galaxy mass, log$(M/M_{\odot})= 10.479\pm0.013$, at which the disk-to-spheroid dominance in their host galaxy sets in is reasonably matched by these N-body + hydrodynamical simulations of gas rich mergers \citep{Stew09a} near the characteristic stellar mass of the galaxy SMF. With increasing host galaxy mass past the crossover threshold, the increase in the spheroid contribution is steep so that galaxies with log$(M/M_{\odot}) \geq$ 11.5 are principally spheroid dominated with $B/T\sim0.9$. The trend levels off at higher stellar masses indicating that even these massive spheroid-dominated, $E$-type galaxies may host disks contributing up to 10\% of their host stellar mass. These spheroid dominant galaxies are likely the ones shown in simulations to have been built up by gas rich major mergers in the early evolutionary history of these objects followed by a series of dry mergers and passive evolution. On the other hand, at lower masses, log$(M/M_{\odot}) \leq$ 10, spheroids still contribute $\sim$30\% of the host mass, and the decline of their fractional contribution is fairly shallow; even at the lowest mass bins above our incompleteness limit, we find that spheroids contribute 10\% or more, which is counter to the conclusion of \citet{Dutt09}. Ongoing numerical simulation projects cited above could refine and test these earlier theoretical predictions against our results.

\indent Similar trends are seen in the galaxy census presented in \S \ref{Nbd2gSM}. At the low end of our stellar mass range, over 75\% of the population fall in the $BT20$ class, with the remaining contributed by the composite classes, in support of the \citet{Dutt09} theoretical results. However, it is also interesting that spheroid-dominated galaxies with $(B/T)_* \geq 0.5$ make up $\sim5\%$ even at the lowest mass end, as shown in the `ying-yang' plot, Figure \ref{Fig12}, lower panel. Canonical theoretical models argue that spheroids are built up principally by gas rich major mergers; these low mass spheroid dominated galaxies may therefore indicate a parallel pathway, akin to monolithic collapse leading to intrinsically spheroidal systems as put forth by \citet{Sale12}. The $blue$ colors of these low stellar mass, spheroidal galaxies, seen in Figure \ref{Fig5}, may be indicative of residual star formation following their initial formation epoch which has not yet been quenched by mergers. 

\indent In Figure \ref{Fig12}, the distributions of the disky, $BT20$ and the pure spheroid, $BT100$ classes peak at either ends of the stellar mass range as expected. However, for the intermediate classes, the distribution for the composite disks peaks at $M \sim 10^{9.5}\,M_{\odot}$, well below the crossover stellar mass, whereas the composite spheroids peak above the crossover point at $M \sim 10^{10.9}\,M_{\odot}$. The physical processes driving these peaks in the number distributions of the composite galaxies as a function of stellar mass are unclear. In our analysis, we have segregated the galaxies only by their $(B/T)_*$ values, and have not taken any effects of environment on their mass make up. Therefore, our results may miss important additional effects due to the environment; Bluck et al. (\textit{in prep}) aim to address this issue by estimating the $(B/T)_*$ evolution of centrals and satellites in an upcoming publication. Simultaneously, it will be interesting to seek matching trends in cosmological simulations of galaxy evolution and thus unravel the cause(s). 

\section {Summary} \label{Sum}

We have presented the SMF and stellar mass density of galaxies, and of their disk and spheroid components in the low redshift universe (median z$\sim$0.1) using a homogeneous sample of 603,122 galaxies selected from the SDSS DR7 spectroscopic catalogs. The structural and photometric properties of this galaxy sample, as well as of their disk and spheroidal components are available in the public release of the GIM2D catalogs \citep{Sima11}. In addition, well characterized stellar masses of these galaxies, as well as those of their two principal building blocks, spheroids and disks, are also available in the public \citet{Mend14} catalogs. These stellar masses have been obtained by SED fitting to five color photometry from SDSS. The combined rich dataset in these catalogs permits us to obtain the SMF and stellar mass densities of the galaxies, as well as those of the spheroidal and disk components. Supported by the statistical significance of our results based on the large sample size, as well as the imaging depth and the sky coverage of the SDSS, our aim is to provide well characterized observational constraints on the properties of the low redshift galaxy population. 

\indent We compare our galaxy SMF and stellar mass densities against those from three other recent observational results from the GAMA survey \citep{Kelv14a}, the PRIMUS \citep{Mous13} and a SDSS-DR7 sample \citep{Bern13}. Our results are largely consistent with these studies in the stellar mass range, $8.9 \leq log(M/M_{\odot}) \leq 11$ from which we draw our principal inferences. We discuss possible reasons for the discrepancies noticed in this comparison especially at the high mass end. Our overall aim is to provide observational constraints against which theoretical results from the ongoing impressive numerical simulations such as the \textit{Eagles} \citep{Scha15}, \textit{Illustris} \citep{Voge14a} and the \textit{FIRE} \citep{Hopk14} projects may be directly compared, and thus advance our understanding of the complex processes driving stellar mass assembly in galaxies. In addition, our results can also be used as a stringent $z\sim0$ baseline for galaxy evolution studies at higher redshifts, such as \textit{CANDELS} \citep{Grog11}, \textit{COSMOS} \citep{Scov07}, and \textit{GOODS} \citep{Dick03} surveys.  

\indent Other than providing the galaxy SMF in the low redshift universe with this sample from SDSS-DR7, the additional take-away results from our efforts are:
\begin{enumerate}[(1)]
\item The stellar mass $(B/T)_*$ ratio of galaxies along with their stellar masses provide a well defined, quantifiable discriminant for their various observed and estimated characteristics. 
\item Based on this, the bivariate rest-frame $(g-r)$ color and stellar mass distribution shows that over 95\% of the spheroid-dominated, $BT100$ galaxies lie along a well-defined \textit{red sequence}, with an equal percentage of their disk-dominated, $BT20$ counterparts restricted to the \textit{blue cloud}. At intermediate $(B/T)_*$ values, galaxies in general do not have intermediate colours, i.e. they lie either on the \textit{red sequence} or in the \textit{blue cloud}. However, since the relative fraction of each changes the most, galaxies in the intermediate $(B/T)_*$ classes appear distributed throughout the \textit{green valley}
\item There is a gradual gradation in color through the \textit{green valley} with increasing $(B/T)_*$ values and stellar mass. Galaxies with higher $(B/T)_*$ ratio migrate from the \textit{blue cloud} to the \textit{red sequence} at lower stellar masses. 
\item Other than the SMF of the galaxies, we also provide the SMF, associated Schechter parameters as well as the stellar mass density of the \textit{spheroidal components} of the entire galaxy sample (not to be confused with the spheroid-dominated galaxy sub-population). Using the corresponding stellar mass in their disk components, we also estimate the SMF, associated Schechter parameters and stellar mass density of the \textit{disk components} of all the galaxies in our sample in the mass range $8.9 \leq log(M/M_{\odot}) \leq 12$. The well characterized GIM2D spheroid+disk decomposition in $SL11$ catalogs, plus associated stellar masses of both components in the $MT14$ catalogs help quantify the statistical and systematic uncertainties associated with our results.  
\item The SMF of the spheroid and disk components permit us to identify the \textit{crossover} stellar mass of log$(M/M_{\odot})$ = 10.3$\pm$0.030 at which they  intersect. We propose that the combination of these two independent SMFs imposes a tighter constraint on any theoretical model aiming to capture all the complex physics and environmental processes which drive the build up of stellar mass in these two principal galaxy components.
\item Even though the SMF of the disk components shows a steep increase at the faint end, their increased volume corrected number density at these lower stellar masses does not compensate for the sharp drop-off seen at the high mass end. Therefore the stellar mass density of the disk components ($0.910\pm0.029$) accounts for only 37\% of the overall galaxy stellar mass density ($2.670\pm0.110$), with the bulk provided by the spheroidal components ($1.687\pm0.063 $), all in units of [$10^8 M_{\odot}\;Mpc^{-3}$]. 
\item Binned by their $(B/T)_*$ ratios, the spheroid-dominated and disk-dominated galaxy sub-populations show similar trends in both their SMFs as well as stellar mass densities as the spheroid and disk components. Specifically, the spheroid-dominated, $(B/T)_* \geq 0.5$ class provides over 65\% of the overall galaxy stellar mass density in the local Universe.
\item The relative contributions of the disk and spheroid stellar masses to that of their host galaxy shows a smooth variation with increasing stellar mass - on average, in a galaxy with $M\sim10^9\;M_{\odot}$, 90\% is in the disk component, while the same fractional contribution comes from the spheroidal component at the high mass end, $M\geq10^{11}\;M_{\odot}$.
\item Lying at the intersection of this trace of relative contributions, a galaxy of stellar mass, $log(M/M_{\odot})\;=\;10.479\pm0.013$, has equal fractional contributions from the spheroid and disk stellar masses.
\item In our census of galaxy number counts, at the low end of the stellar mass range of our sample, $log(M/M_{\odot})\;=\;8.9$, disk-dominated galaxies are predominant ($\geq 85$\%) while spheroid-dominated galaxies rule at the high mass end ($\sim80$\%). The number counts of the intermediate classes show distinct peaks in the intermediate masses, though the mechanisms responsible for these peaks are unclear.  
\item At the highest stellar masses, there are virtually no disk-dominated galaxies. On the other hand however, even at the lowest stellar mass bins, the fractional number of spheroid-dominated galaxies remains $\geq 5$\%, pointing to mechanisms other than mergers for their formation, e.g., monolithic collapse or violent disk instability. On the other hand, we admit that even with our corrections for false disks, a fraction of these galaxies may be artifacts due to GIM2D fitting a fixed n=4 bulge to all galaxies. 
\item We find that the single Schechter function fit adequately describes the underlying SMF in the stellar mass range $8.9 \leq log(M/M_{\odot}) \leq 11.2$ from which we draw our inferences. However, the goodness of fit metrics degrade appreciably at higher stellar masses, especially for the pure spheroid, $BT100$ class, and the disk-dominated $BT20$ sub-population, over- and under-estimating the SMFs respectively. Using a double Schechter function to capture the faint end slope still does not alleviate this issue.  
\item The traces of the Schechter parameters and stellar mass densities of the galaxy sub-populations as functions of their $(B/T)_*$ values shows a near linear variation in all of them. We are using the bivariate distribution of the $(B/T)_*$ values and the stellar masses of our galaxy sample to describe this variation, and aim to publish our findings in a forthcoming paper (Thanjavur et al., \textit{in prep}). 
\end{enumerate}

\indent Taken together our results aim to comprehensively describe the observed low redshift endpoint of galaxy evolution, and thus provide stringent constraints for the large hydrodynamical simulations of galaxy formation, as well as ongoing observational surveys aimed at tracing galaxy evolution to emulate.

\section*{Acknowledgements}
\noindent The results reported here are based on SDSS DR7 photometric and spectroscopic catalogs. The SDSS is managed by the Astrophysical Research Consortium for the Participating Institutions. The Participating Institutions are the American Museum of Natural History, Astrophysical Institute Potsdam, University of Basel, University of Cambridge, Case Western Reserve University, University of Chicago, Drexel University, Fermilab, the Institute for Advanced Study, the Japan Participation Group, Johns Hopkins University, the Joint Institute for Nuclear Astrophysics, the Kavli Institute for Particle Astrophysics and Cosmology, the Korean Scientist Group, the Chinese Academy of Sciences (LAMOST), Los Alamos National Laboratory, the Max-Planck-Institute for Astronomy (MPIA), the Max-Planck-Institute for Astrophysics (MPA), New Mexico State University, Ohio State University, University of Pittsburgh, University of Portsmouth, Princeton University, the United States Naval Observatory, and the University of Washington. Funding for the SDSS and SDSS-II has been provided by the Alfred P. Sloan Foundation, the Participating Institutions, the National Science Foundation, the U.S. Department of Energy, the National Aeronautics and Space Administration, the Japanese Monbukagakusho, the Max Planck Society, and the Higher Education Funding Council for England. The SDSS Web Site is http://www.sdss.org/. 

\bibliographystyle{mnras}
\bibliography{BulgeDisc_SMF_mnras} 

\appendix

\section {Effect of Large Scale Structure on SMF} \label{LSS}

\begin{figure} 
\centering
\includegraphics[width=0.99\columnwidth]{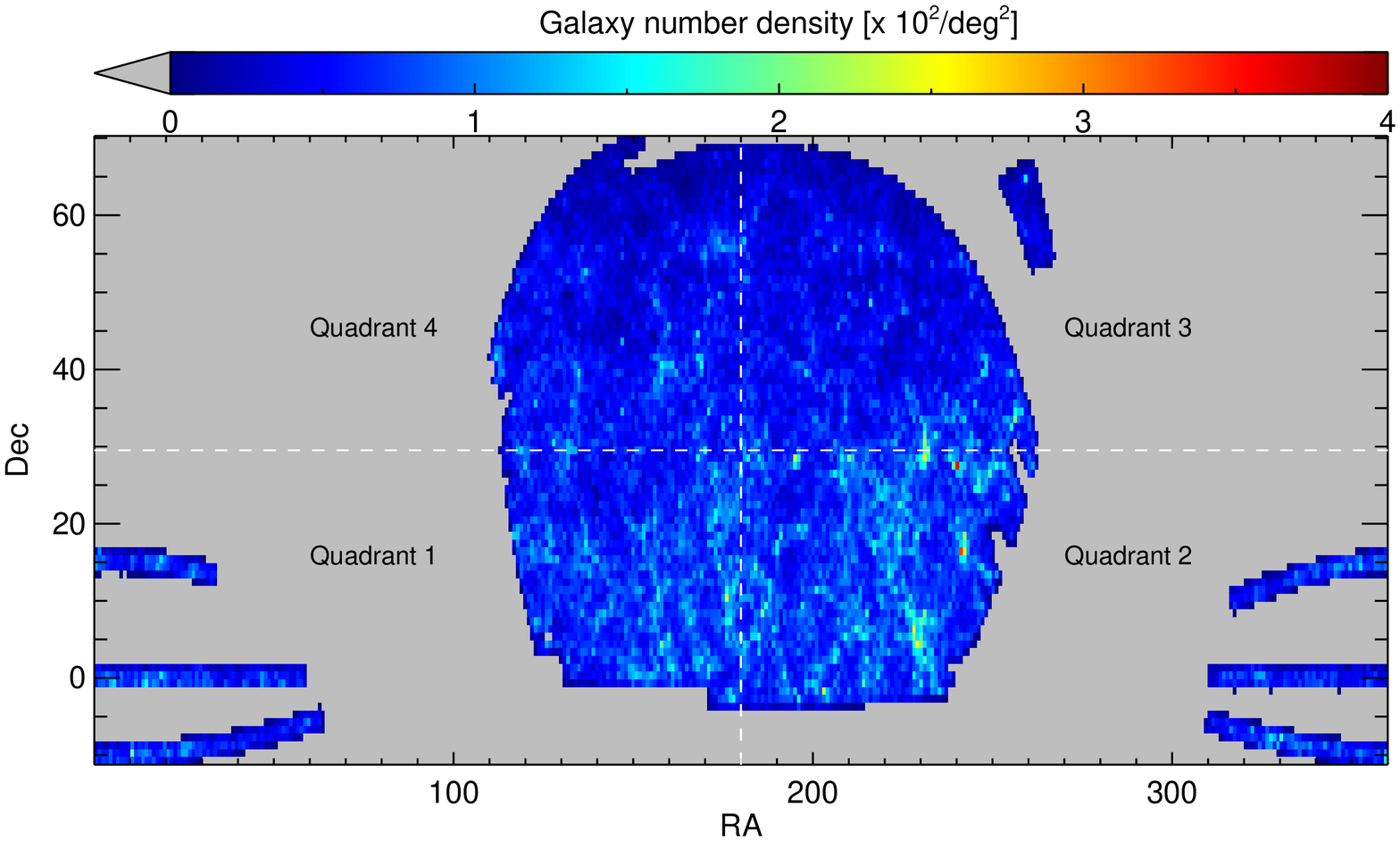}
\includegraphics[width=0.99\columnwidth]{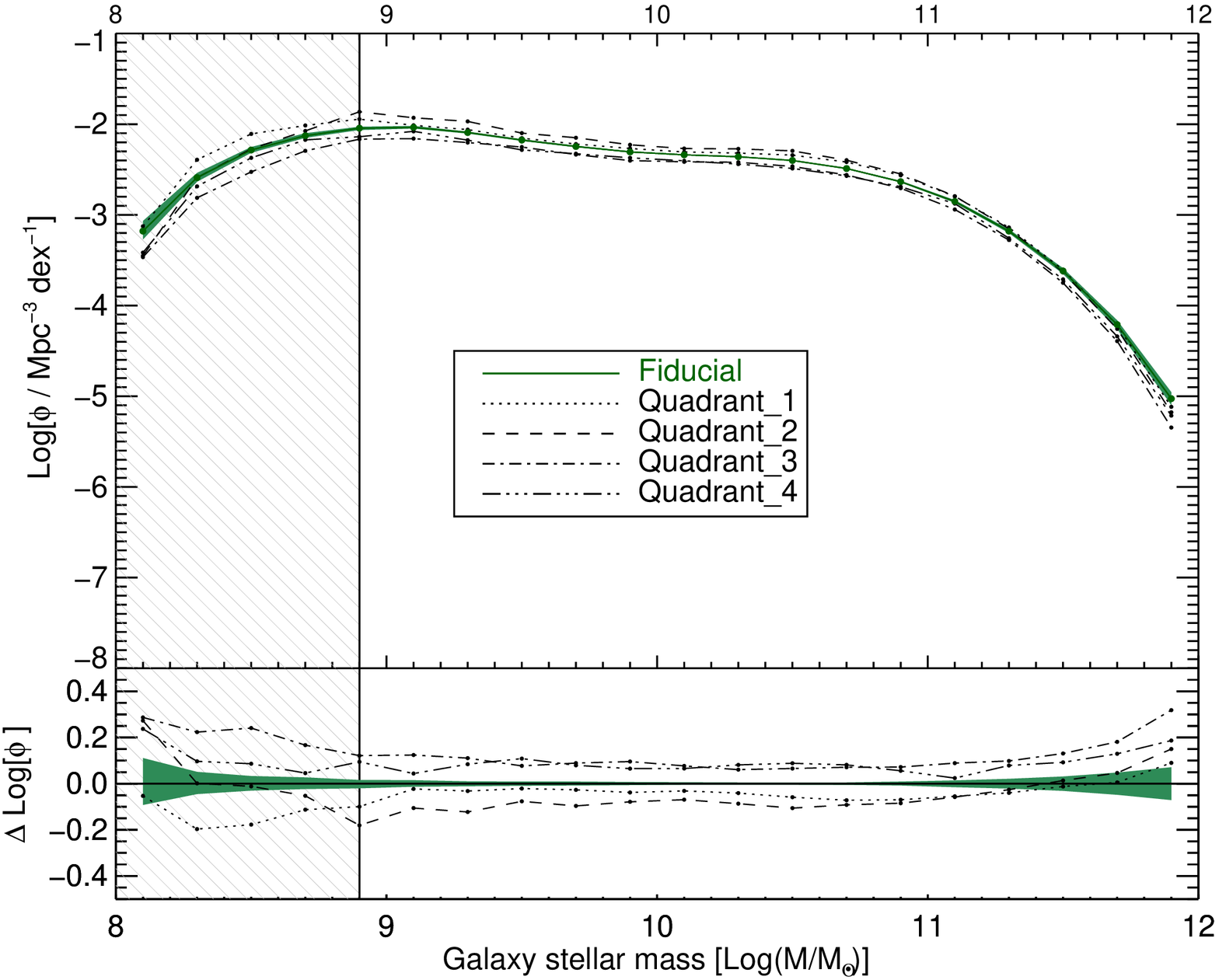}
\caption{(\textit{Top}) Sky positions of the galaxy sample shown as a 2D density distribution to highlight the presence of the peaks due to large scale structure seen in Quadrant 2. (\textit{Bottom}) Comparison of the SMFs of the galaxy subsets in the four quadrants to the fiducial SMF of the complete sample. (\textit{Color figure available in the online journal}) \label{FigA1}}
\end{figure}

\citet{Bald08} have argued that galaxy SMFs obtained using the $1/V_{max}$ method are affected by the variations due to large scale structure (LSS) within the survey region. For the computation of the SMF using the SDSS-DR4 NYU-VACG catalogs containing 49968 galaxies \citep{Blan05b}, they therefore apply an additional weight obtained with the normalized number density for each galaxy. Given the increased survey region SDSS-DR7 and much larger sample size, we expect the local effects of LSS to be smoothed out in the SMF we present. In order to test this, we plot the sky positions of all the galaxies in our sample as a 2D density distribution in the upper panel of Figure \ref{FigA1}. We divide the survey region into four equal quadrants each containing 162241, 204492, 123670 and 112719 galaxies respectively. In the lower panel of Figure \ref{FigA1}, we overplot the SMFs of these four subsets of galaxies on the fiducial SMF of the complete galaxy sample. The enhanced galaxy numbers in Quadrant 2 and to a lesser extent in Quadrant 1 due to LSS lead to no significant differences in the estimated SMFs as shown by their differences in the lower plot of the bottom panel, which are all consistent within their combined uncertainties; for clarity, only the combined systematic and statistical uncertainty of the fiducial SMF is indicated by the green shaded region. Given this consistency between the SMFs even in the presence of LSS in some region of the survey area, we do not include any additional weighting for the local number density while computing the $1/V_{max}$ weighted SMF.

\section{Galaxy SMF internal consistency checks: Contributions from stellar mass fits} \label{IntcompSMFg}

\begin{figure} 
\includegraphics[scale=0.4]{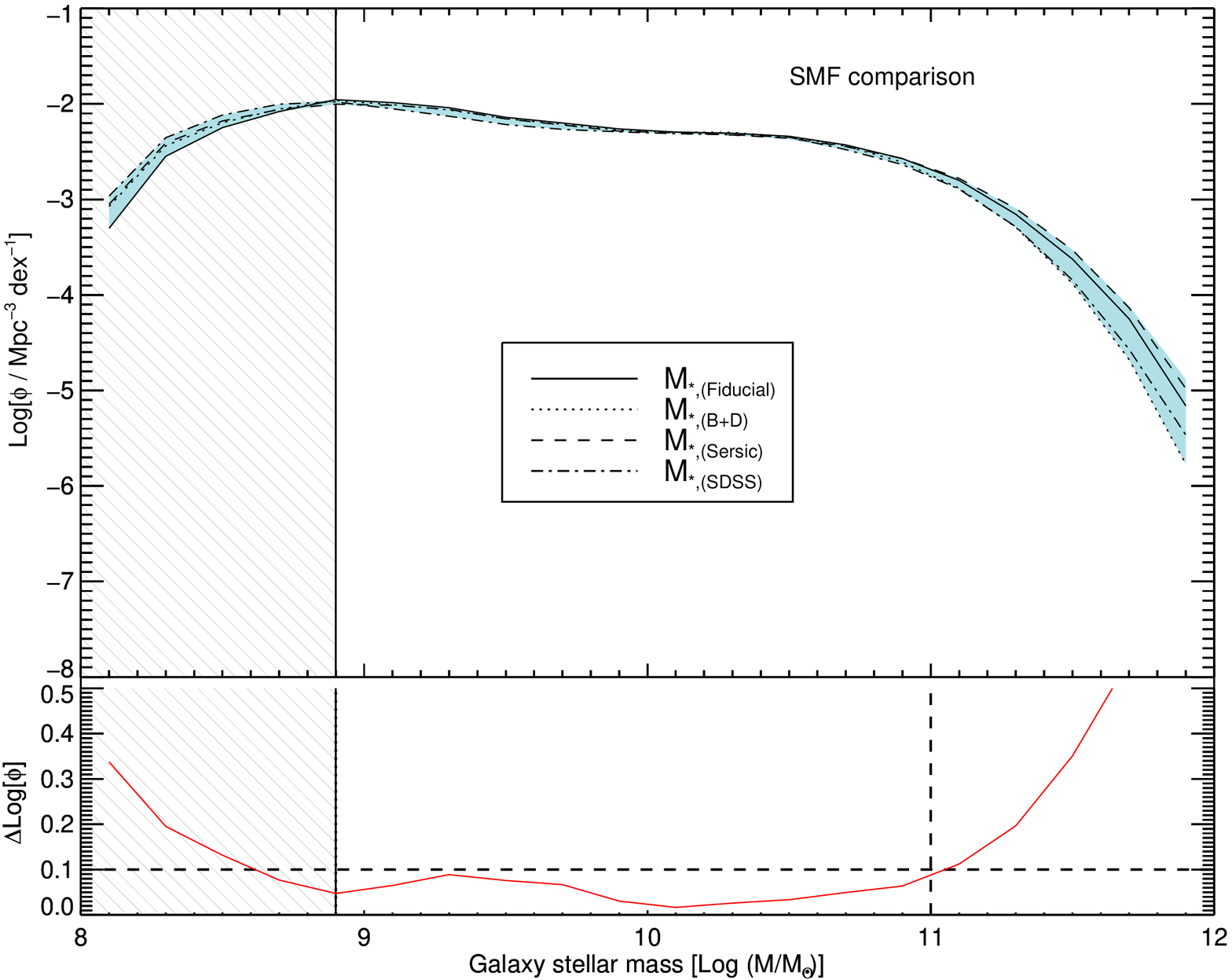}
\caption{Internal consistency checks to assess systematic uncertainties arising from the different fits to the galaxy light profiles used by GIM2D. The GIM2D $M_{*,(b+d)}$ and $M_{*,(bd)}$ stellar masses (see text for description) have systematic offsets $\leq 0.1$dex from a single S\'{e}rsic fit, $M_{*,Sersic}$ and with the stellar masses obtained directly from the SDSS-VAGC catalog for the principal mass range of interest, $8.9 \leq M^* \leq 11.1$. The upper panel shows the comparison of the SMFs, while the lower panel shows the difference between the maximum and minimum values at each stellar mass bin. (\textit{Color figure available in the online journal}) \label{FigB1}}
\end{figure}

The systematic errors in the galaxy stellar masses in the $MT14$ catalogs need to be accounted for in calculating the corresponding systematic and consequently the overall error budget in our SMF estimates, as mentioned in \S \ref{SMFbd}. Given the size of our SDSS-DR7 galaxy sample, the statistical errors in the central stellar mass range, 8.9$ \leq \mathrm{log}(M/M_{\odot}) \leq $11.2 from which we draw our inferences, are at least an order of magnitude smaller than the contributions from the systematics.  $MT14$ discuss in detail the sources of such systematics in stellar masses derived from SED fitting based on the photometric and structural properties of the de Vaucouleurs \citep{deVau48} and exponential light profiles provided by GIM2D. $MT14$ assess the overall uncertainty in stellar masses to lie between 0.1 to 0.2 dex for the bulk of the sample. 

\indent In order to capture the uncertainties in our stellar mass functions arising merely from the different \textit{light profile} fits used to obtain the galaxy stellar masses, in Figure \ref{FigB1} we plot the SMFs obtained independently for each fitting method. For this internal consistency check, we have used the SMF corresponding to three different fitting methods in the $MT14$ catalogs, namely (i) the de Vaucouleur spheroid and exponential disk fitted independently, $M_{*,(b+d)} $ used for the bulk of the galaxies in our work, (ii) the de Vaucouleurs \citep{deVau48} and exponential profiles fitted simultaneously, $M_{*,(bd)}$, and (iii) a single S\'{e}rsic fit, $M_{*,Sersic}$; we have used the latter two fits for correcting for false disks (see \S \ref{BDclasses} for summary details of the corrections, and $MT14$ for a complete description of the fitting methods). In addition, for this comparison we have also included the galaxy stellar masses taken directly from the SDSS-DR7 VAGC catalogs. As a further measure of the systematic uncertainties, in the following Appendix \ref{SystErrChksSMFg}, we extend this to a comparison of the galaxy SMFs derived using a set of 11 different galaxy stellar masses provided in $MT14$, each of which has been obtained with a different set of assumptions in the SED fitting.  

\indent The upper panel of Figure \ref{FigB1} shows a comparison of the galaxy SMFs estimated with the stellar masses obtained using each fitting method. Overplotted is the galaxy SMF presented in this paper obtained by combining the stellar masses from the three different fitting methods based on the selection criteria described in \S \ref{BDclasses}. The spread in the SMFs, taken to be a measure of the systematic uncertainty in the SMF, is highlighted by the colour filled region. However, we do not account for any formal systematic uncertainty in the stellar mass densities from these fitting methods. Figure \ref{FigB1} (\textit{lower}) traces this systematic uncertainty as a function of stellar mass, the spread being computed as the logarithmic difference between the maximum and minimum values at each stellar mass interval. Since the systematic uncertainty from the fitting methods is not expected to be gaussian, we prefer to use the peak-to-peak differences instead of the interquartile values. 

\indent The traces show that the systematic uncertainties are $\leq$0.1 dex in the region $8.9 \leq M^* \leq 11.2$, within which our principal inferences such as the \textit{crossover stellar mass} lie. However, the uncertainties grow quite rapidly on either side of this region, especially toward the higher stellar mass end. This systematic uncertainty needs to be taken into account when using our results, or for comparisons of our results with other similar analyses.

\section{Galaxy SMF internal consistency checks: Contributions from SED fit parameters} \label{SystErrChksSMFg}

\begin{figure} 
\includegraphics[scale=0.4]{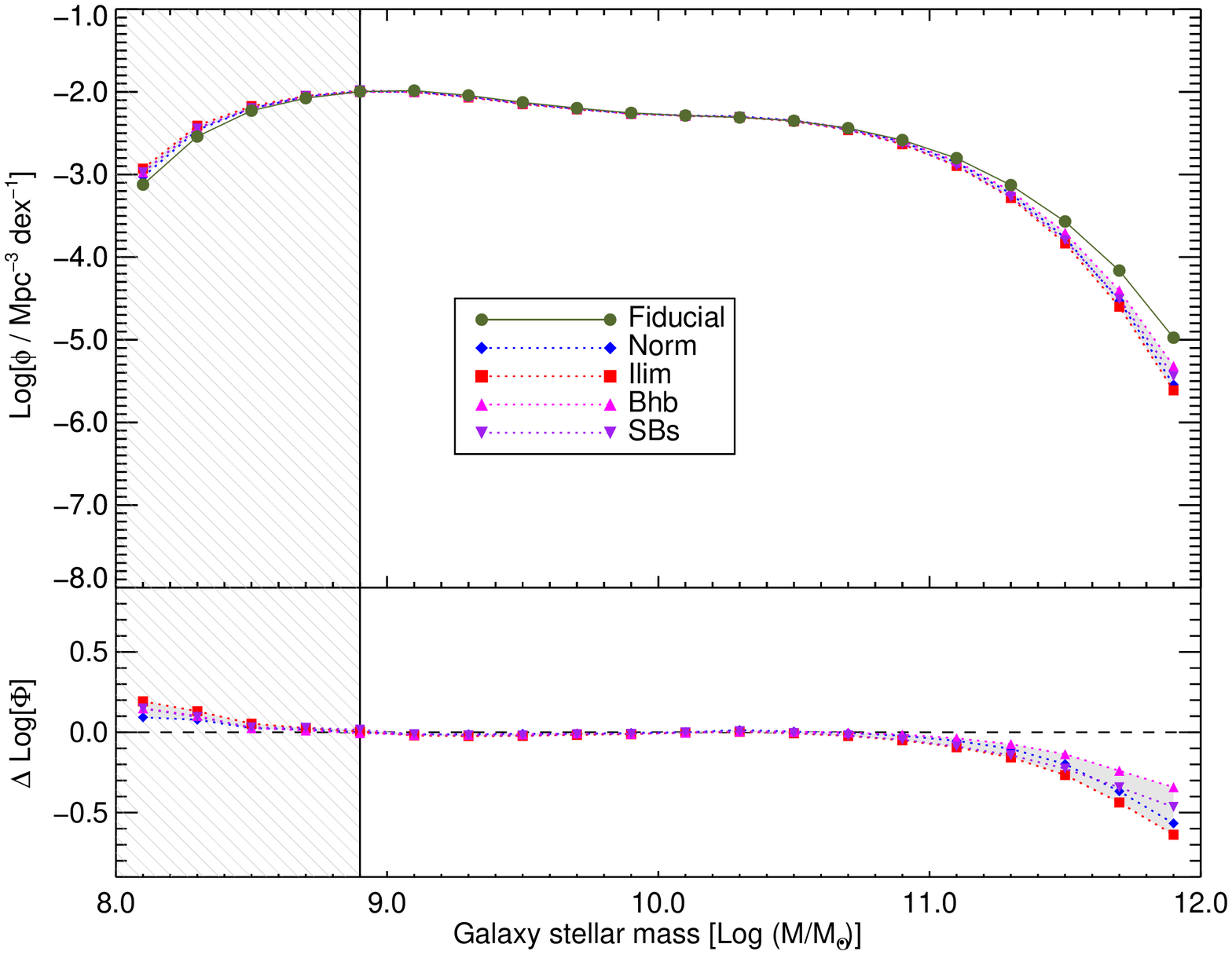}
\caption{Assessment of systematic errors in the galaxy stellar mass function due to assumptions inherent in the SED fitting parameters used for the $MT14$ stellar mass estimates. The galaxy SMF estimated with the fiducial parameter values is compared against the SMF corresponding to a significant change in one FSPS parameter taken in turn, with all others held at fiducial values(see text and $MT14$ for details). (\textit{Upper}) Comparison of the SMFs (\textit{Lower}) Difference between the fiducial and the comparison SMF in dex. Shown in this figure are the effects of excluding 5\% metal poor stars ($Norm$), varying the IMF higher and lower integration limits ($Ilim$), neglecting contribution of blue horizontal branch stars ($Bhb$), and increasing specific frequency of blue stragglers ($SBs$). (\textit{Color figure available in the online journal}) \label{FigC1}}
\end{figure}

\begin{figure} 
\includegraphics[scale=0.4]{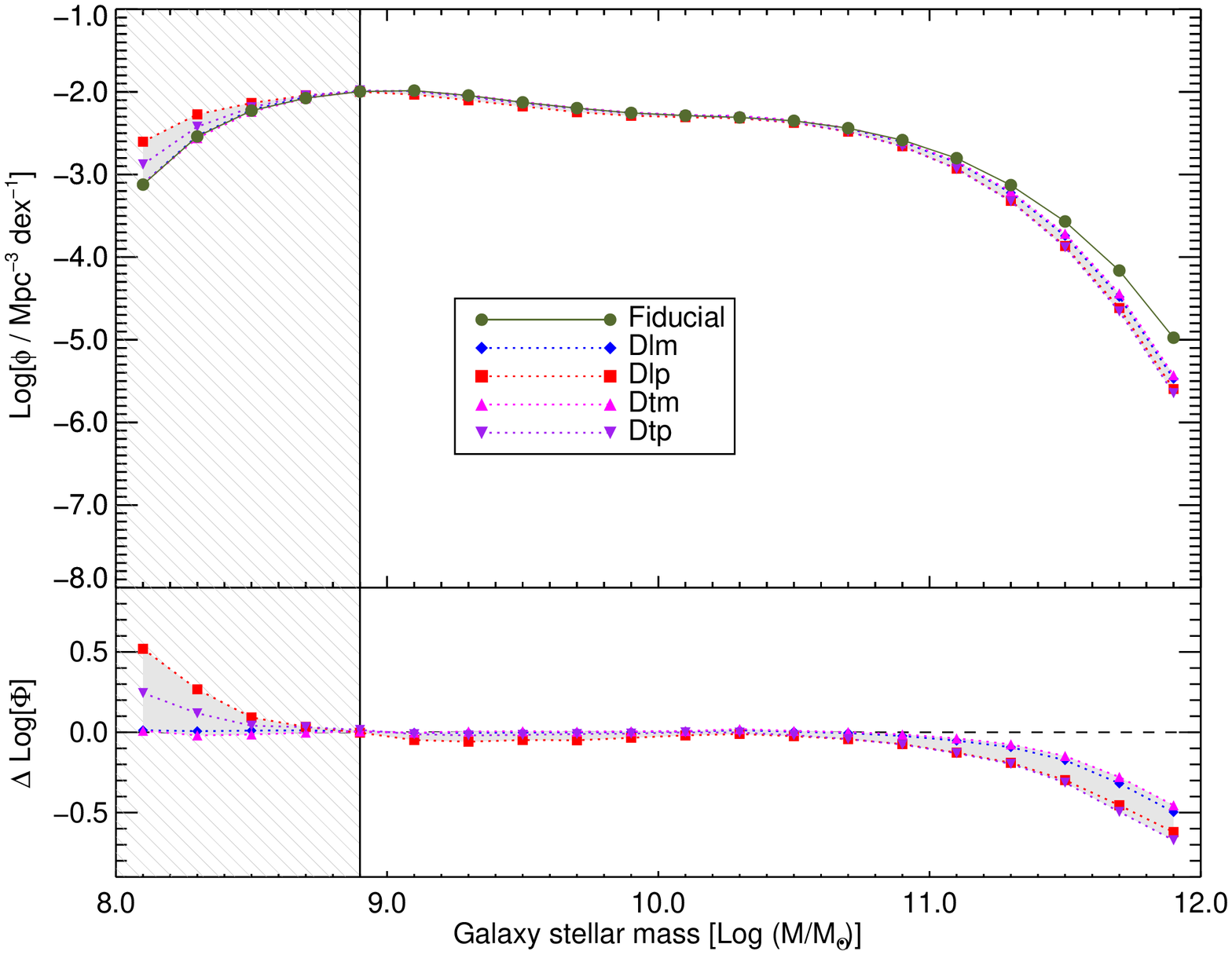}
\caption{Shown in a format similar to Figure \ref{FigC1} are the effects of decreasing ($dlm$) or increasing ($dlp$) the luminosity of thermally pulsating AGB stars by 0.4 dex, as well as decreasing ($dtm$) and increasing ($dtp$) their temperature by 0.2dex. (\textit{Color figure available in the online journal}) \label{FigC2}}
\end{figure}

\begin{figure} 
\includegraphics[scale=0.4]{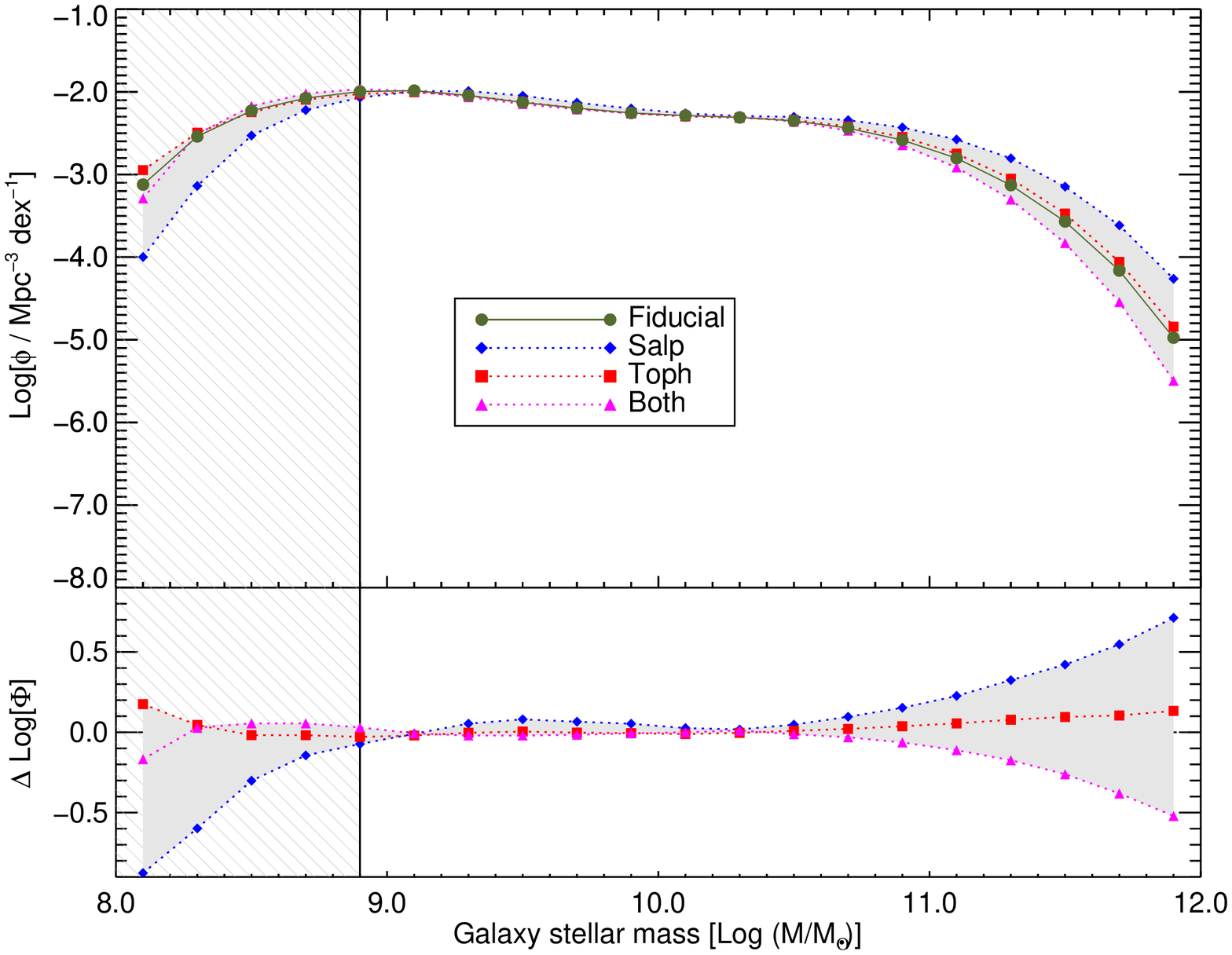}
\caption{Shown in a format similar to Figure \ref{FigC1} are the effects of a Salpeter IMF ($Salp$), increasing the relative number of high mass stars ($Toph$), and third, decreasing the relative numbers of high mass stars while increasing the low mass stars numbers ($Both$). (\textit{Color figure available in the online journal}) \label{FigC3}}
\end{figure}

\begin{figure*} 
\includegraphics[scale=0.75]{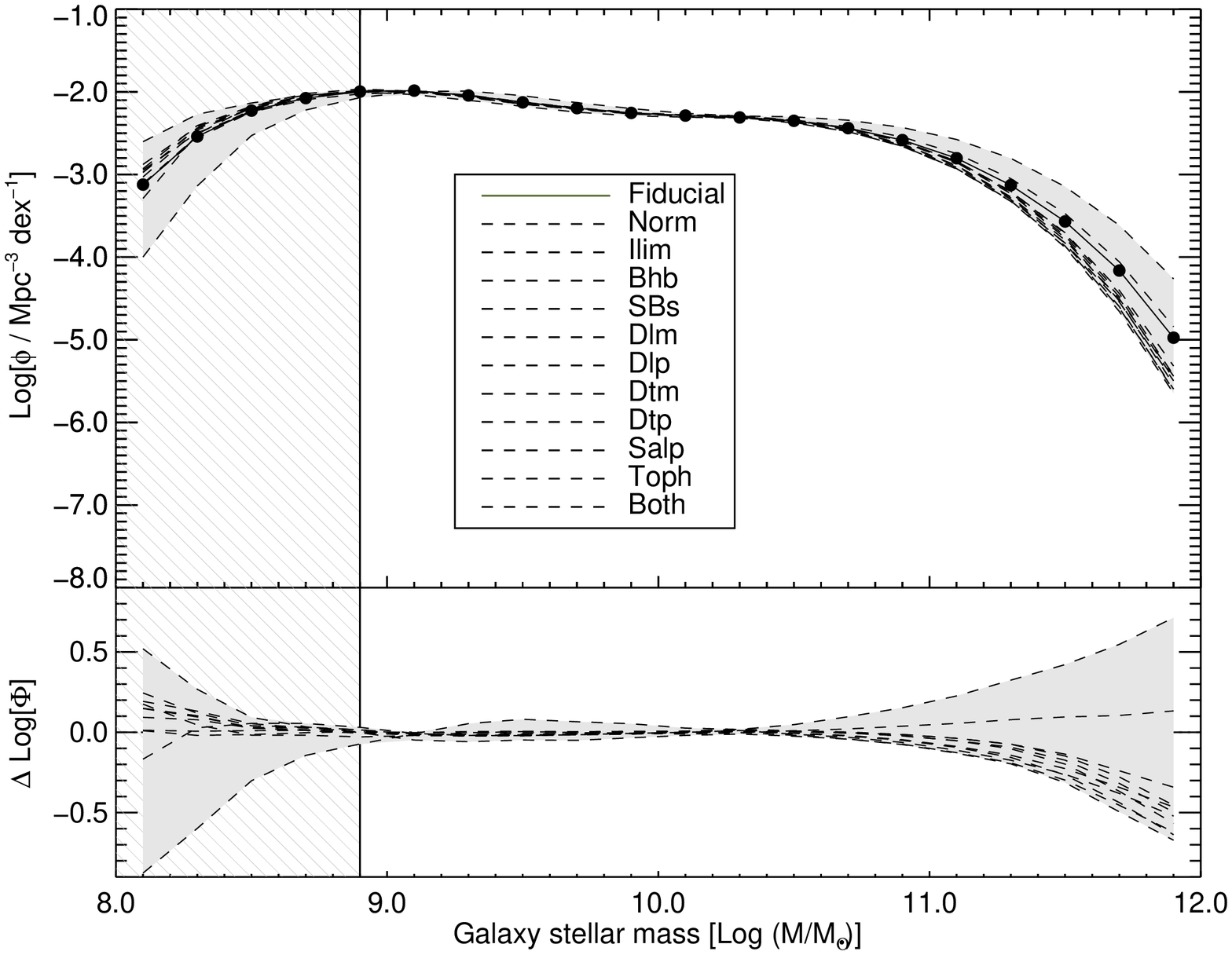}
\caption{Shown in a format similar to Figure {FigC1} are the effects of varying all eleven FSPS parameters independently to highlight the their overall, at times conteracting effects on the SMF. (\textit{Color figure available in the online journal}) \label{FigC4}}
\end{figure*}

\indent Here we aim to assess how much the systematic uncertainties in the stellar masses due to assumptions made in the SED fitting contribute to the galaxy stellar mass functions presented in our paper. We repeat the comparisons performed in Appendix \ref{IntcompSMFg} but this time using 11 different stellar mass estimates provided in an internal release of the $MT14$ catalogs. Each of these is the total bulge+disk stellar mass, $M_{(b+d)}$, but with one SED fitting parameter varied at a time. $MT14$ provide more complete discussions of the effects of each parameter on the resulting stellar masses. The SED fitting parameter which has been varied to obtain the various stellar masses used for this comparison are listed below (the nomenclature has been adopted from $MT14$, and is also used in the set of Figures \ref{FigC1} - \ref{FigC3}:

\textit{Norm} - same as $M_{(b+d)}$, but excluding the $5\%$ metal-poor fraction included in the default modelling

\textit{Ilim} - changing the integration limits of the IMF from 120 $M_{\odot}$ to 100 $M_{\odot}$

\textit{Bhb} - including $20\%$ contribution from blue horizontal branch stars (effectively shifting $20\%$ of the RHB population to be BHB instead)

\textit{Sbs} - increasing the specific frequency of blue stragglers to 2 (relative to the number of HB stars)

\textit{Dlm} - decreasing the luminosity of tp-AGB stars by 0.4 dex

\textit{Dlp} - same as above, except increasing the luminosity of tp-AGB stars by 0.4 dex

\textit{Dtm} - decreasing the temperature of tp-AGB stars by 0.2 dex

\textit{Dtp} - same as above, but increasing the temperature of tp-AGB stars by 0.2 dex

\textit{Salp} - masses computed using a Salpeter IMF \citep{Salp55} instead of Chabrier \citep{Chab03}.

\textit{Toph} - changing the high-mass slope of the IMF from 1.3 to 1.0 (for $M> 1M_{\odot}$), increasing the relative number of high-mass stars

\textit{Both} - changing the high-mass slope of the IMF from 1.3 to 1.6, while also increasing the relative number of low-mass stars.\\

Each of the Figures \ref{FigC1} - \ref{FigC3} compares the galaxy stellar mass functions estimated with one set of the variant stellar masses with the fiducial stellar mass we have obtained using the selection criteria described in \S \ref{BDclasses}. The set of stellar masses being used in each comparison is shown annotated. The upper panel in each figure compares the stellar mass functions; the lower panel traces the logarithmic difference between the SMFs as a function of galaxy stellar mass. Finally,  Figure \ref{FigC4} we present a compilation of the SMFs corresponding to all the parameters overplotted in order to visually highlight their overall, and at times contradictory effects on the systematic uncertainties of the galaxy SMF. 

\indent Overall, the combined effect of all these parameters in the central stellar mass range 8.9$ \leq log(M/M_{\odot}) \leq $11.2 which forms the focus of our work is $\leq 0.05$dex. However, at the lower stellar mass range, $log(M/M_{\odot}) <$8.9, the choice of the SED fitting parameters is seen to lead to significantly different trends in the SMF. At the same time, our SDSS-DR7 galaxy sample also suffers from increasing incompleteness in this stellar mass range, and hence we have excluded this range from our analysis. Similar trends are seen in the high stellar mass range, $log(M/M_{\odot}) >$11.2, where the low space density of galaxies and the diverging trends due to the choice of parameters also lead to significant, and at times contradictory, systematic uncertainties in the SMF. Even though we draw our inferences mainly based on trends in the central stellar mass ranges, we caution the reader about these higher systematics at either end of the stellar mass range of our study.

\section{SMF values of galaxies, spheroids and disks} \label{SMFapp}

 \begin{table*}
 \caption{Stellar Mass Functions of galaxies, disks and spheroids}
 \begin{center}
 \scriptsize
 {\renewcommand{\arraystretch}{2.0}
 \begin{tabular}{|C{3cm}|C{3cm}|C{3cm}|C{3cm}|}
 \hline
 Stellar mass & All galaxies & Disks only & Spheroids only \\
 \hline
 log($M$) & log($\phi$) & log($\phi$) & log($\phi$) \\
 \hline
  [$M_{\odot}$] & [$Mpc^{-3}\;dex^{-1}$] & [$Mpc^{-3}\;dex^{-1}$] & [$Mpc^{-3}\;dex^{-1}$] \\
\hline
8.100 & $-3.129^{+0.111}_{-0.092}$ & $-2.625^{+0.056}_{-0.055}$ & $-2.227^{+0.045}_{-0.047}$ \\ 
8.300 & $-2.539^{+0.051}_{-0.045}$ & $-2.305^{+0.036}_{-0.038}$ & $-2.190^{+0.031}_{-0.029}$ \\ 
8.500 & $-2.235^{+0.033}_{-0.030}$ & $-2.127^{+0.024}_{-0.022}$ & $-2.180^{+0.016}_{-0.017}$ \\ 
8.700 & $-2.077^{+0.027}_{-0.023}$ & $-2.012^{+0.018}_{-0.018}$ & $-2.191^{+0.012}_{-0.017}$ \\ 
8.900 & $-1.994^{+0.016}_{-0.020}$ & $-1.959^{+0.013}_{-0.014}$ & $-2.225^{+0.010}_{-0.013}$ \\ 
9.100 & $-1.985^{+0.015}_{-0.013}$ & $-2.003^{+0.008}_{-0.010}$ & $-2.281^{+0.013}_{-0.013}$ \\ 
9.300 & $-2.042^{+0.010}_{-0.011}$ & $-2.084^{+0.007}_{-0.008}$ & $-2.349^{+0.010}_{-0.007}$ \\ 
9.500 & $-2.125^{+0.009}_{-0.009}$ & $-2.160^{+0.007}_{-0.007}$ & $-2.416^{+0.007}_{-0.007}$ \\ 
9.700 & $-2.196^{+0.009}_{-0.008}$ & $-2.225^{+0.007}_{-0.008}$ & $-2.457^{+0.007}_{-0.006}$ \\ 
9.900 & $-2.256^{+0.007}_{-0.006}$ & $-2.290^{+0.007}_{-0.004}$ & $-2.473^{+0.005}_{-0.007}$ \\ 
10.100 & $-2.288^{+0.006}_{-0.004}$ & $-2.359^{+0.005}_{-0.004}$ & $-2.475^{+0.004}_{-0.004}$ \\ 
10.300 & $-2.309^{+0.005}_{-0.004}$ & $-2.471^{+0.003}_{-0.003}$ & $-2.480^{+0.003}_{-0.004}$ \\ 
10.500 & $-2.350^{+0.005}_{-0.004}$ & $-2.655^{+0.005}_{-0.004}$ & $-2.523^{+0.004}_{-0.004}$ \\ 
10.700 & $-2.439^{+0.005}_{-0.005}$ & $-2.928^{+0.008}_{-0.007}$ & $-2.620^{+0.006}_{-0.006}$ \\ 
10.900 & $-2.584^{+0.009}_{-0.008}$ & $-3.308^{+0.022}_{-0.010}$ & $-2.778^{+0.010}_{-0.009}$ \\ 
11.100 & $-2.802^{+0.014}_{-0.015}$ & $-3.787^{+0.015}_{-0.015}$ & $-3.002^{+0.014}_{-0.014}$ \\ 
11.300 & $-3.129^{+0.021}_{-0.021}$ & $-4.361^{+0.024}_{-0.024}$ & $-3.305^{+0.021}_{-0.021}$ \\ 
11.500 & $-3.569^{+0.031}_{-0.031}$ & $-5.062^{+0.039}_{-0.039}$ & $-3.710^{+0.031}_{-0.031}$ \\ 
11.700 & $-4.161^{+0.048}_{-0.048}$ & $-5.955^{+0.089}_{-0.084}$ & $-4.268^{+0.049}_{-0.048}$ \\ 
11.900 & $-4.977^{+0.071}_{-0.071}$ & $-6.782^{+0.241}_{-0.372}$ & $-5.049^{+0.071}_{-0.071}$ \\ 
\hline
\end{tabular}
}
 \end{center}
 \begin{flushleft}
 {\footnotesize Stellar mass functions of galaxies, spheroids and disks in the low redshift universe ($z\sim0.1$) estimated using over 600,000 galaxies drawn from SDSS-DR7, using $S11$ spheroid/disk structural and photometric properties, and with corresponding stellar masses estimated with SED fitting to five color photometry in $MT14$. This table and the table with the stellar mass functions of the four galaxy sub-populations (discussed in \S \ref{SMFbddom}) are also available in machine readable format as online data accompanying this publication.}
\end{flushleft}
\label{gbdSMFvals}
 \end{table*}
 \normalsize


\bsp	
\label{lastpage}
\end{document}